\documentstyle[epsf,aps]{revtex}
\begin{document}

\input amssym.def                    

\newcommand{\identity}[0]{{\Bbb I}}

\begin{titlepage}
\title{Static and Dynamic Properties of Dissipative Particle Dynamics}
\author{C. Marsh\\
Theoretical Physics, Oxford University\\
1 Keble Road, Oxford OX1 3NP UK\\
and\\ 
G. Backx and M.H. Ernst\\
Institute for Theoretical Physics, Utrecht University\\
Princetonplein 5, P.O. Box 80.006, 3508 TA Utrecht, The Netherlands}

\date{February 4, 1997}
\label{lns}
\maketitle
\begin{center}
\section*{Abstract}
\end{center}

The algorithm for the DPD fluid, the dynamics of which is
conceptually a combination of 
molecular dynamics, Brownian dynamics and lattice gas automata, is designed for
simulating rheological properties of complex fluids on hydrodynamic
time scales. This paper calculates the equilibrium and transport
properties (viscosity, self-diffusion) of the thermostated DPD fluid explicitly in 
terms of the system parameters. It is demonstrated that
temperature gradients cannot exist, and that there is therefore
no heat conductivity.

Starting from the $N$-particle Fokker-Planck, or Kramers' equation,
we prove an $H$-theorem for the free energy, obtain 
hydrodynamic equations, and derive a nonlinear kinetic equation (the 
Fokker-Planck-Boltzmann equation) for the single particle distribution function. 
This kinetic equation is solved by the Chapman-Enskog method. The analytic results 
are compared with numerical simulations.

\vspace{2cm}

Keywords: Dissipative Dynamics, Fokker-Planck Equation, Transport Coefficients, 
Kinetic Theory, Computer Simulation Techniques.

\end{titlepage}

\section{Introduction}
\label{section1}
The interest of the last decade in dynamical and rheological
properties of complex fluids has seen the introduction of
several new numerical methods
for carrying out computer simulations on hydrodynamic time scales, the
simulation of which using molecular dynamics often results in
intensive computational demands. These new techniques include: (i) lattice gas 
cellular automata (LGCA) \onlinecite{Doolen,RZ};
(ii) lattice Boltzmann equation (LBE) \onlinecite{Succi} and (iii) dissipative 
particle dynamics (DPD).

The last method was introduced by Hoogerbrugge and Koelman \onlinecite{HK:DPD}, 
and was modified by Espa\~{n}ol and Warren 
\onlinecite{PEPW:DPD} to ensure a proper thermal equilibrium
state. The primary goal of this paper is a theoretical 
analysis and explicit calculation of transport and 
thermodynamic properties in terms of model parameters.
This is highly relevant in view of the current interest in
applications of DPD to systems such as flows past complex objects 
\onlinecite{HK:DPD}, concentrated colloidal suspensions \onlinecite{KH:POLY,BCL}, 
dilute polymer solutions \onlinecite{SHM:POLY,Madden} and phase separation 
\onlinecite{CN:DPD}.

The DPD algorithm models a fluid of $N$ interacting particles out of
equilibrium and conserves mass and momentum. Position and velocity
variables are {\em continuous}, as in Molecular Dynamics (MD), but time is updated 
in {\em discrete} steps $\delta t$, as in LGCA and LBE. The algorithm is a mixture 
of molecular dynamics, Brownian and Stokesian dynamics and LGCA's, with a {\em 
collision} and a
{\em propagation} step. In the collision step each particle interacts with all the 
particles inside an action sphere of radius
$R_0$ through {\em conservative} forces ${\bf F}_{ij}$, {\em
dissipative} forces ${\bf F}_{D,ij}$, which are proportional to both
the stepsize $\delta t$ and a friction constant $\gamma$, and {\em
random} forces ${\bf F}_{R,ij}$, which supply the energy lost by the
damping. Here $i,j \in \left\{ 1,2,...N \right\}$ label the
particles. In numerical simulations, this is implemented by
simultaneously updating the velocities from their precollision value
${\bf v}_{i}$ to their postcollision value ${\bf v}_{i}^{*}$ according
to the instantaneous forces exerted by all particles inside the action
sphere. In the subsequent {\em propagation} step of fixed length $\delta t$ all 
particles move freely to their new positions ${\bf r}_i(t+\delta 
t)={\bf r}_i(t)+{\bf v}_i^{*} \delta t$.

Usual forms of the conservative force mean that the particles may be
considered as completely interpenetrable.
These softer interactions have the computational
advantage \onlinecite{HK:DPD,SHM:POLY} of allowing particle motion on
the order of a mean free path $l_0$ during each time step of fixed
length $\delta t$.  This represents a substantial advantage over event driven MD 
algorithms for hard sphere fluids, where the length $\delta t$ of the free 
propagation interval is on average much shorter, especially at fluid densities.

By ignoring some of the microscopic details of the interactions, which are 
presumably irrelevant for fluid dynamics, DPD has the advantages of LGCAs, but 
avoids the disadvantages of lacking Galilean invariance and of introducing 
spurious conservation laws.
In fact, the ``point particles'' should not be considered as molecules in a fluid, 
but rather as clusters of particles that interact dissipatively 
\onlinecite{HK:DPD,PEPW:DPD}. The introduction of noise and dissipation represents 
a coarse-grained mesoscopic level of description and hydrodynamic behaviour is 
expected at much smaller particle numbers than in conventional MD. If $t_0 \simeq 
1/\gamma n R_0^d$ denotes the characteristic kinetic timescale in DPD, with 
$n=N/V$ the number density, $d$ the number of dimensions and $\gamma$ the friction 
constant, then $t_0$ is considered to be large compared to any molecular time 
scale.

In this coarse-grained description the dominant interactions are the
dissipative and random forces, whereas the conservative forces can be
interpreted as weak forces of relatively long range and may be taken
into account as a Vlasov mean field term in the kinetic equations. In
addition, they can have the spurious effect of tending to force the DPD
particles into ``colloidal crystal'' configurations, unless friction
and noise are sufficiently large to prevent cooling into a lattice
configuration \onlinecite{HK:DPD,PEPW:DPD,VDB}. In the second half of
the paper, where we derive a kinetic equation for the single particle
distribution function, the conservative force will be neglected. This
corresponds to the strong damping limit ($\gamma$ large). The random
forces act effectively as repulsive forces to prevent collapse of DPD particles.

A substantial contribution towards the understanding of the DPD fluid
was given by Espa\~{n}ol and Warren \onlinecite{PEPW:DPD}, who derived
a Fokker-Planck equation for the $N$-particle distribution function in
the limit of continuous time ($\delta t \rightarrow 0$). These authors
also modified the original algorithm by imposing the detailed balance
conditions, which guarantee the existence of the proper thermal Gibbs'
equilibrium, described by $\exp[-H/\theta_0]$ where $H$ is the
Hamiltonian of the corresponding conservative system 
and $\theta_0=k_B T_0$ is the global equilibrium
temperature. These results are briefly reviewed in Section
\ref{section2} to establish the notation. Concerning the macroscopic
evolution equations, Espa\~{n}ol formally established
\onlinecite{PE:HD} the linearized Navier-Stokes equations and derived
Green-Kubo formulae for the DPD transport coefficients using a Mori-Zwanzig 
projection operator technique. However, to date no quantitative evaluation of 
these formulae for DPD seems to
exist. Hence, little is known explicitly about the approach to equilibrium, 
the validity of standard hydrodynamics (system size dependence, effects of 
generalised hydrodynamics) or about transport coefficients. 
For the transport coefficients, Hoogerbrugge and Koelman
\onlinecite{HK:DPD} have estimated the kinematic viscosity
$\nu=\eta/\rho$, where $\eta$ is the shear viscosity and $\rho=n m$
is the mass density, as $\nu \sim \gamma n R_0^{d+2}$ with $n R_0^d \sim
1$. This result has recently been extended in \onlinecite{ME1,VDB} to include the 
bulk viscosity by applying the ``continuum approximation'' to the discrete 
equations of motion for the DPD particles, following suggestions of Hoogerbrugge 
and Koelman \onlinecite{HK:DPD}.
In Section \ref{section3} we show how the free energy of the DPD fluid
{\em monotonically} approaches its equilibrium value by proving an $H$-theorem for 
the Fokker-Planck equation of Espa\~{n}ol and Warren, and we make the connection 
with the detailed balance conditions derived in \onlinecite{PEPW:DPD}.

As a first step towards establishing the full nonlinear hydrodynamic
equations we derive in Section \ref{section4} the full macroscopic
conservation laws for mass and momentum density, as well as the energy
balance equation (details are given in Appendix A). The conceptual
basis for the existence of hydrodynamic equations is the {\em local
equilibrium} state, which in DPD is very different from that in a molecular
fluid, because of the unusual role of the temperature. In Section
\ref{section5} we study in a quantitative fashion the decay of the
energy density $e({\bf r},t)$ and ``kinetic'' temperature $\theta({\bf
r}, t)$ towards thermal equilibrium with global temperature
$\theta_0$, and we assess in what sense and on what timescale the DPD
fluid describes an {\em isothermal} fluid out of equilibrium. This is
done on the basis of a nonlinear kinetic equation -referred to as
Fokker-Planck-Boltzmann (FPB) equation- for the single particle
distribution function $f({\bf x}, t)$. It will be obtained from the
first equation of the BBGKY hierarchy for the DPD fluid in combination with the 
molecular chaos assumption.

By solving in Section \ref{section6} the FPB equation in the
hydrodynamic stage, using the Chapman-Enskog method, we derive the
constitutive relations and the Navier-Stokes equation. This enables us
to calculate in Section \ref{section7} the transport coefficients of
shear and bulk viscosity, as well as the self-diffusion coefficient.

So far, we have not discussed the {\em discrete time version} of DPD, as 
implemented in actual simulations. They show a sensitive dependence of
thermodynamic and transport properties on time step $\delta t$ 
\onlinecite{ME1,VDB,MY}.
A promising step towards understanding the $\delta t$-dependence was recently 
taken by Marsh and Yeomans \onlinecite{MY}, who calculated the equilibrium 
temperature as a function of the step size, determined stability criteria 
for the step size, and validated their result by extensive numerical 
simulations. We shall not attempt to present here a systematic study of the 
different ${\cal O}(\delta t)$-corrections to equilibrium distributions and 
transport properties, but postpone this to a later publication. 

The paper ends in Section \ref{section8} with comments on the most important 
results and future prospects for DPD.

\section{The Fokker-Planck Formalism}
\label{section2}

The dynamics of a DPD system defines the time evolution of an $N$-particle
system, specified by a point $\Gamma = \{ {\bf x}_i =({\bf v}_i,{\bf r}_i)
\vert i=1,2 \cdots N \}$ in phase space, in terms of stochastic
differential equations.  For a theoretical description it is more convenient
to consider the equivalent Fokker-Planck equation, derived by Espa\~{n}ol
and Warren \onlinecite{PEPW:DPD}.

To interpret the separate terms in the Fokker-Planck equation, it is
instructive first to consider the analogous {\em Kramers' equation} for
the probability $P({\bf v},{\bf r},t)$ of a single particle of mass $m$, having a
phase description ${\bf x}=({\bf v},{\bf r})$ at time $t$:

\begin{equation} \label{eqn:KRAMERS}
\partial_t P  + {\bf v} \cdot \frac{\partial}{\partial {\bf r}} P = -
\frac{{\bf F}({\bf r})}{m} \cdot \frac{\partial}{\partial {\bf v}} P
+\gamma \frac{\partial}{\partial {\bf v}} \cdot {\bf v}P + \frac{\sigma^{2}}{2}
\frac{\partial^{2}}{\partial {\bf v}^{2}} P.
\end{equation}
The three terms on the right can be interpreted as follows.  The first term
is an external conservative force ${\bf F}({\bf r}) = -\nabla V(r)$.
The term involving the damping constant $\gamma$ corresponds to the
Langevin  force $-\gamma {\bf v}$ and the diffusive term with diffusion
coefficient $\frac{1}{2} \sigma^{2}$ results from the random force
$\sigma {\bf \hat{\xi}}$ in the equivalent Langevin description, which reads:

\begin{eqnarray} \label{eqn:LANGEVIN}
\frac{d{\bf r}}{dt} &=& {\bf v} \nonumber \\
\frac{d{\bf v}}{dt} &=& \frac{{\bf F}}{m} - \gamma {\bf v} +
\sigma {\bf \hat{\xi}},
\end{eqnarray}
where $\sigma{\bf \hat{\xi}}$ is Gaussian white noise with amplitude
$\sigma$ and $< {\bf \hat{\xi}} >=0$ and
$< {\bf \hat{\xi}}(t) {\bf \hat{\xi}}(t') > = \identity \delta(t-t')$, where 
$\identity$ is a $d$-dimensional unit tensor.

If we impose that the stationary solution of the Kramers' equation be the Gibbs' 
distribution:  $P_{eq} \sim \exp \{ - (\frac{1}{2} m{\bf v}^{
2}+V(r))/\theta_0 \}$, then the diffusion
coefficient must satisfy the following {\em Detailed Balance} (DB) condition:

\begin{equation} \label{eqn:DB}
\sigma^{2} = \frac{2\gamma \theta_0}{m},
\end{equation}
where $\theta_0=k_B T_0$ is the temperature in thermal equilibrium, measured
in energy units.

The full Fokker-Planck equation derived by Espa\~{n}ol and Warren for
the DPD system is a direct extension of the Kramers' 
equation to $N$ interacting particles.  The time evolution
of the $N$-particle distribution function $P({\bf \Gamma},t)$ is governed by:

\begin{equation} \label{eqn:FP}
\partial_t P = \left( {\cal L}_{\rm C} + {\cal L}_{\rm D} + {\cal L}_{\rm R} 
\right) P,
\end{equation}
where the Conservative, Dissipative and Random parts of the evolution
operator are defined respectively as:

\begin{eqnarray} \label{eqn:OPERATORS}
{\cal L}_{\rm C} &=& -\sum_{i} \left( {\bf v}_i \cdot  \frac{\partial}{\partial 
{\bf r}_i}
+\frac{{\bf F}_i}{m}\cdot \frac{\partial}{\partial {\bf v}_i} \right) 
\nonumber \\
&=& -\sum_{i} {\bf v}_i\cdot \frac{\partial}{\partial {\bf r}_i}
-\frac{1}{2} \sum_{i,j \neq i} \frac{{\bf F}({\bf R}_{ij})}{m}\cdot 
\left( \frac{\partial}{\partial {\bf v}_i}-
\frac{\partial}{\partial {\bf v}_j} \right) \nonumber \\
{\cal L}_{\rm D} &=& \sum_{i,j \neq i} \gamma w_D(R_{ij})
\left( {\bf \hat{R}}_{ij}\cdot \frac{\partial}{\partial {\bf v}_i} \right)
\left( {\bf \hat{R}}_{ij}\cdot {\bf v}_{ij} \right) \nonumber \\
{\cal L}_{\rm R} &=& \sum_{i,j \neq i} \frac{\sigma^{2}}{2} w_R^2(R_{ij})
\left( {\bf \hat{R}}_{ij}\cdot \frac{\partial}{\partial {\bf v}_i} \right)
{\bf \hat{R}}_{ij}\cdot \left( \frac{\partial}{\partial {\bf v}_i}-
\frac{\partial}{\partial {\bf v}_j} \right).
\end{eqnarray}
The summations run over all particles and the only difference to the
original \onlinecite{PEPW:DPD} is that the parameters $\gamma$ and $\sigma$ have
been scaled by the mass $m$ such that $\gamma$ has the dimensions of an inverse 
time.  The three terms above are the
$N$-particle extensions of the three terms on the rhs of (\ref{eqn:KRAMERS}).

\begin{itemize}
\item
The conservative part ${\cal L}_{\rm C}$ results from the additive and central
interparticle interactions due to a potential, $V=\frac{1}{2}
\sum_{i,j \neq i} \phi(R_{ij})$ where ${\bf R}_{ij}={\bf r}_i-{\bf r}_j$
is the relative position and a hat denotes a unit vector.  It is the
Liouville operator for the corresponding conservative system and in the limit of 
zero noise 
and friction, equation (\ref{eqn:OPERATORS}) 
reduces to the Liouville equation.
\item
The second term is analogous to the dissipative term in the Kramers' equation.
It accounts for the Langevin damping force between
the pair $(ij)$, which is proportional to the friction constant $\gamma$ and
to the component of the relvative velocity ${\bf v}_{ij}$ along the
line of centres ${\bf \hat{R}}_{ij}$, and is of finite range.  This last
property is described by a positive weighting function $w_D(R_{ij})$ that is
only non-vanishing inside an action sphere of finite radius $R_0$.
\item
The last term in (\ref{eqn:OPERATORS}) represents the random noise and
should be compared with the diffusive term in (\ref{eqn:KRAMERS}).  The
random force $\sigma \hat{\xi}_{ij}$ between the pair $(ij)$ is directed along
${\bf \hat{R}}_{ij}$ and is proportional to $\sigma w_R(R_{ij})$ where the
weighting function $w_R(R_{ij})$ is again only non-vanishing within a finite
action sphere.
\end{itemize}
The ranges of the conservative, dissipative and random forces may all
be different as the model stands.
Moreover, one of the essential properties of DPD is that its dynamics
conserves total particle number, $N$, and the total momentum, 
${\bf P}=\sum_i m {\bf v}_i$.  Consequently $<N>=\int d{\bf x} f({\bf x},t)$
and $<P> = \int d{\bf x} m{\bf v} f({\bf x},t)$ are constants of the motion.
The latter is always set equal to 0 as the total system is assumed to be
at rest. Here $f({\bf x},t)$ is the single particle distribution function.

In addition, we want to emphasize that microscopic momentum
conservation is an essential property of a fluid model if it is to
have a momentum density $\rho({\bf r},t) {\bf u}({\bf r},t)$ that is
slowly varying in space and time.  In contrast, we note that the
energy of the system is not strictly conserved under the DPD algorithm.
The equations for the mass, momentum and energy density will be discussed
more fully in Section \ref{section4}. 
In typical applications the conservative force may be set equal to zero.
The parameters $\gamma$ and $\sigma$ satisfy the equation
(\ref{eqn:DB}) and the weighting functions are chosen such that
$w_D(r)=w_R^2(r)$, which constitute in combination with (\ref{eqn:DB})
the detailed balance conditions for the DPD system, as will be discussed in 
Section \ref{section3}.
The density is typically chosen such that there are 5 to 10 particles
within an action sphere, which means that the instantaneous total force
on any particle is small on average.

Figure 1 shows an enlargement of part of configuration space, showing
a sequence of 20 consecutive particle positions evolving from a randomly chosen
initial configuration.  The trajectories are relatively smooth, in contrast
to the discontinuous paths in hard core interactions, illustrating
that the resultant force on each particle is relatively small at this
parameter setting.

\begin{figure}[h]
\epsfxsize 12cm
\epsfysize 10cm
\epsfbox{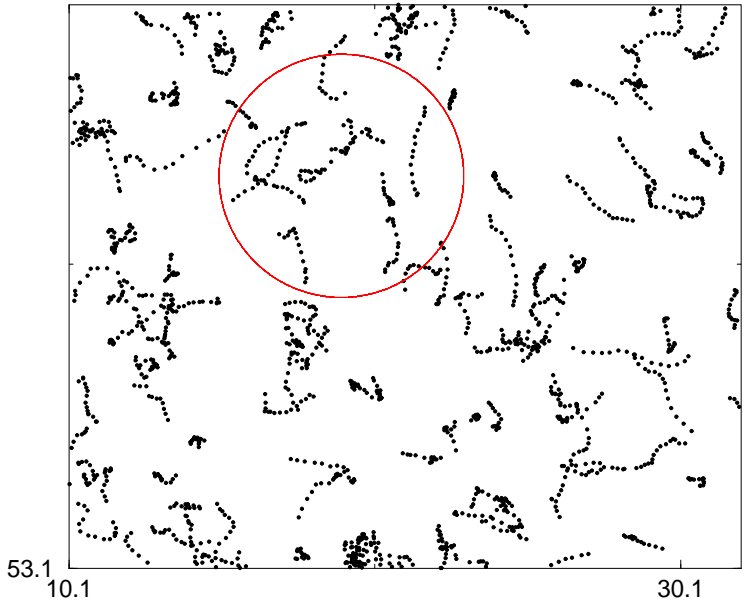}
\caption{\label{fig1} Typical evolution of a particle configuration over a period 
of $20$ timesteps, showing that DPD interactions are ``soft'' as compared to hard 
core interparticle interactions. The circle with radius $R_0=4$ indicates the 
range of interaction.}
\end{figure}

\section{An H-Theorem}
\label{section3}

Consider the following functional of the $N$-particle distribution function
$P({\bf \Gamma},t)$:

\begin{equation} \label{eqn:FUNCTIONAL}
{\cal F} [P] = \int d{\bf \Gamma} P({\bf \Gamma},t) \left\{
H+\theta_0 \ln P({\bf \Gamma},t) \right\},
\end{equation}
where $\theta_0=m\sigma^{2}/2\gamma$ and $H$ is the Hamiltonian of the
corresponding conservative system:

\begin{equation} \label{eqn:HAMILTONIAN}
H = \sum_i \frac{1}{2}m{\bf v}_i^2 + V = 
\sum_i \frac{1}{2}m{\bf v}_i^2 + \frac{1}{2} \sum_{i, j \neq i}
\phi(R_{ij}),
\end{equation}
$V$ is the potential energy and $\phi(R_{ij})$ is the pair interaction.
The functional can be interpreted as a sort of
{\em free energy} ${\cal F}=E-\theta_0 S$, where $E=<H>$ is the average
total energy and $S=-<\ln P>$ yields the total entropy.  The goal of this
section is to show that $\cal F$ is a Lyapunov functional with
$\partial_t {\cal F} \leq 0$ and to investigate the implications of this
result for the equilibrium solution of the Fokker-Planck equation.

The time derivative of (\ref{eqn:FUNCTIONAL}) yields in combination with
(\ref{eqn:FP}):

\begin{equation} \label{eqn:F_EVOL}
\partial_t {\cal F} = \int d{\bf \Gamma} \left\{
H+\theta_0 \ln P + \theta_0 \right\} \left(
{\cal L}_{\rm C} + {\cal L}_{\rm D} + {\cal L}_{\rm R} \right) P({\bf \Gamma},t).
\end{equation}
We observe that the third term inside $\{ \cdots \}$ in (\ref{eqn:F_EVOL})
vanishes due to total probability conservation. Then consider the
contribution $\{\partial_t {\cal F}\}_{\rm C}$ to (\ref{eqn:F_EVOL}) due to the
Liouville operator ${\cal L}_{\rm C}$.  Partial integration with respect to
${\bf r}_i$ and ${\bf v}_i$ yields directly
\begin{equation} \label{eqn:F_EVOL_C}
\{\partial_t {\cal F}\}_{\rm C} = -\int d{\bf \Gamma} \left\{
P {\cal L}_{\rm C} H + \theta_0 {\cal L}_{\rm C} P \right\}.
\end{equation}
Here ${\cal L}_{\rm C} H = \{ H,H \} = 0$ because the curly brackets represent 
Poisson
brackets, as can easily be demonstrated.  The second term in
(\ref{eqn:F_EVOL_C}) reduces to surface terms in ${\bf r}_i$ and
${\bf v}_i$ and therefore vanishes too.

Next, we combine the remaining terms in (\ref{eqn:F_EVOL}) and perform
a partial ${\bf v}_i$ integration, also symmetrizing the result over
$i$ and $j$. The final result is:

\begin{eqnarray} \label{eqn:F_EVOL_FULL}
\partial_t {\cal F} &=& -\frac{1}{2} m\gamma \int d{\bf \Gamma}
\sum_{i,j \neq i} \left\{ {\bf \hat{R}}_{ij}\cdot {\bf v}_{ij}
+\frac{\theta_0}{m} {\bf \hat{R}}_{ij}\cdot \left(
\frac{\partial}{\partial {\bf v}_i}-\frac{\partial}{\partial {\bf v}_j}
\right) \ln P \right\} \nonumber \\
&& \times \left( w_D(R_{ij}) {\bf \hat{R}}_{ij}\cdot {\bf v}_{ij}
+w_R^2(R_{ij}) \frac{\theta_0}{m} {\bf \hat{R}}_{ij}\cdot \left(
\frac{\partial}{\partial {\bf v}_i}-\frac{\partial}{\partial {\bf v}_j}
\right) \right) P.
\end{eqnarray}
Now we make the following observation.  If we choose:

\begin{equation} \label{eqn:WEIGHTS}
w_D(r) = w_R^2(r) \equiv w(r),
\end{equation}
where $w(R)$ is an arbitrary positive function vanishing for $r > R_0$,
then $\partial_t {\cal F} \leq 0$, as the rhs of (\ref{eqn:F_EVOL_FULL})
can be cast into the form:

\begin{equation} \label{eqn:F_DECREASES}
\partial_t {\cal F}  = -\frac{1}{2} m\gamma \int d{\bf \Gamma} P
\sum_{i,j \neq i} w(R_{ij}) \left\{ \cdots \right\}^{2} \leq 0,
\end{equation}
where $\{\cdots\}$ is the same as in (\ref{eqn:F_EVOL_FULL}). Note that the 
equality sign applies if and only if $P$ is the solution of 
(\ref{eqn:EQM_CONSTRAINT}) below. Consequently, the free energy-type function 
${\cal F}[P]$ is a monotonically decreasing function of time, until it reaches 
equilibrium where $P=P_{eq}$ which is simply the solution of $\{\cdots\}$=0 for 
every pair $(ij)$:

\begin{equation} \label{eqn:EQM_CONSTRAINT}
\left\{ {\bf v}_{ij} + \frac{\theta_0}{m} \left(
\frac{\partial}{\partial {\bf v}_i}-\frac{\partial}{\partial {\bf v}_j}
\right) \right\} P_{eq} = 0.
\end{equation}
Changing variables to the relative velocities of the particles, it is easy
to prove that the equilibrium distribution of the system is separable in
the velocities, and has the general form:

\begin{equation} \label{eqn:GENERAL_EQM}
P_{eq}({\bf \Gamma}) = A({\bf r}_1,\cdots,{\bf r}_N) \exp
\left\{ -\frac{1}{2\theta_0} \sum_i m ( {\bf v}_i - {\bf u}_0)^{2} \right\}\mbox{ 
, }
\end{equation}
where ${\bf u}_0$ is a constant independent of ${\bf r}_i$ and $t$.  We will only
consider macroscopic systems which are not in uniform motion at long times
and consequently limit ourselves to ${\bf u}_0=0$.

The function $P_{eq}({\bf \Gamma})$ is also the stationary solution of the
Fokker-Planck equation (\ref{eqn:FP}) if $A({\bf r}_1,\cdots,{\bf r}_N)$
satisfies ${\cal L}_{\rm C} A = 0$.  This yields the Gibbs' distribution as the
equilibrium solution:

\begin{equation} \label{eqn:GIBBS}
P_{eq}({\bf \Gamma}) = \frac{1}{Z} \exp \left\{ -\frac{H}{\theta_0} \right\},
\end{equation}
where $H$ is the Hamiltonian (\ref{eqn:HAMILTONIAN}) of the system and $Z$
is a normalisation constant. We assume that $P_{eq}$ is uniquely determined by the 
requirements that it satisfies (\ref{eqn:EQM_CONSTRAINT}) and (\ref{eqn:FP}).
Consequently the DPD system will always reach
the same equilibrium state if left undriven, independent of the volume
and number of particles.  The temperature of this equilibrium state has
a value $\theta_0=m\sigma^2/2\gamma$, which {\em only} depends on
the parameters of the model.  So DPD describes a system, thermostated at
$\theta_0$ and with a free energy ${\cal F}[P_{eq}]$ at equilibrium.
Note that, in contrast to the $H$-theorem for the Boltzmann equation
(see e.g. \onlinecite{REICHL}), no molecular chaos approximation is required to 
derive this result for the DPD system.

In their original discussion \onlinecite{PEPW:DPD}, Espa\~{n}ol and Warren
imposed that the Gibbs' distribution be the stationary solution of the
Fokker-Planck equation.  The consequences of this requirement can be
seen by inserting (\ref{eqn:GIBBS}) into (\ref{eqn:FP}).  It leads to the
so-called {\it detailed balance} constraint

\begin{equation} \label{eqn:DB_CONSTRAINT}
w_D(r) = \frac{\sigma^{2}m}{2\gamma\theta_0}w_R^2(r) = w_R^2(r)=w(r).
\end{equation}
Consequently, the constraint imposed in (\ref{eqn:WEIGHTS}) is
the detailed balance constraint for DPD.  The important result from the
$H$-theorem is that it demonstrates that the Gibbs' distribution
(\ref{eqn:GIBBS}) is the {\em inevitable} equilibrium
distribution.
Throughout the rest of the paper we shall restrict ourselves to dealing
exclusively with DPD systems that obey the DB condition
(\ref{eqn:DB_CONSTRAINT}). If the DB condition is violated, no $H$-theorem can be 
derived and the
Gibbs' distribution is not a stationary solution of the FP equation for DPD.
In this case, the stationary state of the system does not correspond to thermal
equilibrium but to some driven state, which will in general exhibit long
range spatial correlations (see e.g.
\onlinecite{GG:NEQ,DKS:CORREL,BE:CORREL}).

The original version of DPD, introduced by Hoogerbrugge and Koelman
\onlinecite{HK:DPD} violates the DB requirement (\ref{eqn:DB_CONSTRAINT}) and 
therefore its
stationary distribution will not approach a Gibbs' state but may exhibit
spatial correlations, e.g. algebraic correlations $r^{-d}$ where $d$ is the
number of dimensions, extending far beyond the ranges of the conservative,
dissipative and stochastic forces.  This absence of thermal equilibrium
is likely to be the reason for difficulties and inconsistencies
discussed in \onlinecite{HK:DPD,PEPW:DPD,PE:HD}.

\section{Macroscopic Conservation and Balance Equations}
\label{section4}

In the previous section we have established the existence of and approach to
a thermal equilibrium state for the DPD system that obeys the detailed
balance condition (\ref{eqn:DB_CONSTRAINT}).  In this section, we address
the problem of how the quantities of macroscopic interest evolve in time
towards the final equilibrium state, concentrating on the local mass density
$\rho({\bf r},t) = mn({\bf r},t)$, the local momentum density
$\rho({\bf r},t) {\bf u}({\bf r},t)$ and the local energy density
$e({\bf r},t)$.

As discussed in section \ref{section2} the microscopic dynamics of DPD
conserve mass and momentum, and the corresponding macroscopic densities
obey local conservation laws.  As the total energy is not conserved under
DPD, the evolution equation for the macroscopic energy density does not
have the form of a local conservation equation, but contains source and
sink terms corresponding to the random and dissipative forces respectively.
In the final equilibrium state, these will balance each other.

Consider a general macroscopic quantity $<A>$, defined through

\begin{equation} \label{eqn:MACRO_DEFN}
\left< A \right> = \int d{\bf \Gamma} A({\bf \Gamma}) P({\bf \Gamma},t).
\end{equation}
Its time evolution can be obtained from the Fokker-Planck equation
(\ref{eqn:FP}) combined with the detailed balance condition
(\ref{eqn:DB_CONSTRAINT}), by multiplying the Fokker-Planck equation with
$A({\bf \Gamma},t)$, integrating over all ${\bf \Gamma}$-space and performing
one or two partial integrations with respect to ${\bf v}_i$ and ${\bf v}_j$.
The result is the general rate of change equation:

\begin{eqnarray} \label{eqn:GRCE}
\partial_t \left< A \right> &=& \left< \sum_i \left( {\bf v}_i\cdot 
\frac{\partial}{\partial {\bf r}_i} + \frac{{\bf F}_i}{m}\cdot 
\frac{\partial}{\partial {\bf v}_i} \right) A \right>
-\gamma \left< \sum_{i, j \neq i} w(R_{ij}) \left\{
{\bf \hat{R}}_{ij}\cdot {\bf v}_{ij} \right\} \left\{ {\bf \hat{R}}_{ij}\cdot 
\frac{\partial}{\partial {\bf v}_i} \right\} A \right> \nonumber \\
&+& \frac{\gamma \theta_0}{m} \left< \sum_{i,j \neq i} w(R_{ij})
\left\{ {\bf \hat{R}}_{ij}\cdot \frac{\partial}{\partial {\bf v}_i} \right\}
\left\{ {\bf \hat{R}}_{ij}\cdot \left(
\frac{\partial}{\partial {\bf v}_i} - \frac{\partial}{\partial {\bf v}_j}
\right) \right\} A \right>
\end{eqnarray}
for any dynamic variable $A({\bf \Gamma})$.

Consider the conserved mass density $\rho({\bf r},t) = mn({\bf r},t)$ and
the momentum density $\rho({\bf r},t) {\bf u}({\bf r},t)$ defined through:

\begin{eqnarray} \label{eqn:FIELD_DEFN}
n({\bf r},t) &=& \left< \sum_i \delta ( {\bf r}-{\bf r}_i ) \right>
= \int d{\bf v} f({\bf v},{\bf r},t) \nonumber \\
n{\bf u}({\bf r},t) &=& \left< \sum_i {\bf v}_i \delta ( {\bf r}-{\bf r}_i )
\right> = \int d{\bf v} f({\bf v},{\bf r},t) {\bf v} .
\end{eqnarray}
It is convenient at this stage to introduce the single particle and
pair distribution functions, defined as:

\begin{eqnarray} \label{eqn:F_DEFN}
f({\bf x},t) &=& f({\bf v},{\bf r},t) = \left< \sum_i
\delta( {\bf x}-{\bf x}_i) \right> \nonumber \\
f^{(2)}({\bf x},{\bf x}',t) &=& \left< \sum_{i,j \neq i} \delta
({\bf x}-{\bf x}_i) \delta ( {\bf x}'-{\bf x}_j ) \right>.
\end{eqnarray}
Application of (\ref{eqn:GRCE}) to the conserved densities in
(\ref{eqn:FIELD_DEFN}) yields the macroscopic conservation laws:

\begin{eqnarray} \label{eqn:MACRO_CONS}
\partial_t \rho &=& -\mbox{\boldmath $\nabla$} \cdot  \rho {\bf u} \nonumber \\
\partial_t (\rho {\bf u}) &=& -\mbox{\boldmath $\nabla$}\cdot ( \rho {\bf uu} + 
{\bf \Pi} ),
\end{eqnarray}
where $\mbox{\boldmath $\nabla$}=\partial/\partial {\bf r}$ and ${\bf \Pi}$ is the 
local pressure
tensor or momentum flux density in the local rest frame of the fluid.
The continuity equation has been derived by setting $A=\sum_i \delta ({\bf r}-{\bf 
r}_i)$ into (\ref{eqn:GRCE}). The only non-vanishing term is the one containing 
$(\partial/\partial {\bf r}_i) \delta({\bf r}-{\bf r}_i)=-\mbox{\boldmath 
$\nabla$} 
\delta({\bf r}-{\bf r}_i)$, and the continuity equation follows at once. 
Derivation of the conservation equation for the momentum density proceeds along 
similar lines by choosing $A=\sum_i m {\bf v}_i \delta({\bf r}-{\bf r}_i)$. 
Details of the latter derivation are given in the Appendix \ref{appendix1}, where 
it is shown that

\begin{equation} \label{eqn:PT_DEFN}
{\bf \Pi} ({\bf r},t) = {\bf \Pi}_{\rm K} ({\bf r},t) + {\bf \Pi}_{\rm C}
({\bf r},t) + {\bf \Pi}_{\rm D} ({\bf r},t),
\end{equation}
with kinetic(K), collisional transfer(C) and dissipative(D) contributions:

\begin{eqnarray} \label{eqn:PT_CONTRIBS}
{\bf \Pi}_{\rm K} &=& \int d{\bf v} \; m {\bf VV} f({\bf v},{\bf r},t)
\nonumber \\
{\bf \Pi}_{\rm C} &=& \frac{1}{2} \int d{\bf v} d{\bf v}' \int d{\bf R} \;
{\bf R}{\bf F}({\bf R})\overline{f}^{(2)}({\bf v},{\bf r},{\bf v}',{\bf r}',t)
\nonumber \\
{\bf \Pi}_{\rm D} &=& -\frac{1}{2} m \gamma \int d{\bf v} d{\bf v}' \int d{\bf 
R}\;
w(R) \left\{ {\bf R}\cdot  ( {\bf v}-{\bf v}' ) \right\} {\bf \hat{R}\hat{R}}
\overline{f}^{(2)}({\bf v},{\bf r},{\bf v}',{\bf r}',t) ,
\end{eqnarray}
with ${\bf R}={\bf r}-{\bf r}'$.
The kinetic flux contains the so-called {\em peculiar velocity}, 
${\bf V} = {\bf v}-{\bf u}({\bf r},t)$, and $\overline{f}^{(2)}$ is the
spatially averaged pair distribution function:

\begin{equation} \label{eqn:F2BAR}
\overline{f}^{(2)} ({\bf v},{\bf r},{\bf v}',{\bf r},t) = \int^1_0 \:
d\lambda f^{(2)} ({\bf v},{\bf r}+\lambda {\bf R},{\bf v}',
{\bf r}+(\lambda-1) {\bf R},t).
\end{equation}
The kinetic and collisional transfer contributions to the momentum flux in
(\ref{eqn:PT_DEFN}) and (\ref{eqn:PT_CONTRIBS}) are present in any particle
model with conservative forces.  They are dominant in systems with
sufficiently high density - dense gases and liquids - where the potential
energy contributions are non-negligible with respect to the kinetic fluxes.

The explicit form for these collisional transfer contributions is given in
the literature for several cases: smooth potentials \onlinecite{RESIBOIS},
elastic hard spheres \onlinecite{CC}, or inelastic hard spheres
\onlinecite{JR}.  The dissipative contribution
${\bf \Pi}_{\rm D}$ results from the Langevin-type damping forces between the
particles.  The random forces do not contribute to the momentum flux.

The $H$-theorem, derived in section \ref{section3}, guarantees the approach
to thermal equilibrium, where the distribution functions take the form:

\begin{eqnarray} \label{eqn:F_EQM}
&f({\bf x})& = n_0 \varphi_0 ({\bf v}) = n_0 \left( \frac{m}{2\pi \theta_0}
\right)^{d/2} \exp \left\{ -\frac{m {\bf v}^{2}}{2\theta_0} \right\}
\nonumber \\
&f^{(2)}({\bf x},{\bf x}')& = n_0^{2} \varphi_0({\bf v}) \varphi_0({\bf v}')
g(\vert {\bf r}-{\bf r}' \vert),
\end{eqnarray}
where $g(R)$ is the pair distribution function in thermal equilibrium,
$n_0=N/V$ is the number density, and $\varphi_0({\bf v})$ the Maxwellian
velocity distribution.

The sum of the kinetic and collisional transfer contributions reduces to the
equilibrium pressure and the dissipative contribution vanishes.  Thus,
${\bf \Pi}=p_0 \identity$ with $p_0$ given by the virial theorem:

\begin{equation} \label{eqn:VIRIAL}
p_0 = n_0\theta_0 -\frac{n_0^{2}}{2d} \int d{\bf R} R \frac{d\phi(R)}{dR} g(R),
\end{equation}
where
\begin{equation} \label{eqn:FORCE_DEFN}
{\bf F}({\bf R}) = -{\bf \hat{R}} \frac{d\phi(R)}{dR}.
\end{equation}

Away from global equilibrium, the pressure tensor ${\bf \Pi}({\bf r},t)$
will contain the local equilibrium pressure and terms involving the
viscosities.  However, before the Navier-Stokes equations, or more
generally, the full set of hydrodynamic equations, can be derived the concept
of {\em local equilibrium} - which forms the conceptual basis of slow
hydrodynamic evolution - has to be re-examined, as the energy is no longer
a conserved quantity.  This can only be done after identifying the slow and fast 
relaxation modes in DPD, on the basis of a kinetic equation.  This will be done in 
section \ref{section5}.

Next we consider the energy density, defined as:

\begin{eqnarray} \label{eqn:E_DENSITY}
e({\bf r},t) &=& \left< \sum_i \epsilon_i({\bf v}) \delta({\bf r}-{\bf r}_i) 
\right>
\nonumber \\
&=& \int d{\bf v} \frac{1}{2} m {\bf v}^2 f({\bf x},t) +
\frac{1}{2} \int d{\bf v} d{\bf v}' \int d{\bf R} \phi(R)
f^{(2)}({\bf v},{\bf r},{\bf v}',{\bf r}-{\bf R},t) \mbox{ , }
\end{eqnarray}
where $\epsilon_i ({\bf v})=\frac{1}{2} m V_i^2 + \sum_{i \neq j} \phi(R_{ij})$ is 
the microscopic energy per particle. 
Use of the rate of change equation (\ref{eqn:GRCE}) leads after some lengthy
algebra to the energy balance equation, as discussed in Appendix
\ref{appendix1}.  It reads:

\begin{equation} \label{eqn:E_FLOW}
\partial_t e = -\mbox{\boldmath $\nabla$} \cdot {\bf q} + \Gamma.
\end{equation}
Here the explicit form of the source term \onlinecite{source} is:

\begin{equation} \label{eqn:SOURCE}
\Gamma({\bf r},t) = \gamma \left< \sum_{i,j \neq i} w(R_{ij})
\left\{ \theta_0 - \frac{m}{2} \left\{ {\bf \hat{R}}_{ij}\cdot \left(
{\bf v}_i-{\bf v}_j \right) \right\}^2 \right\} \delta({\bf r}-{\bf r}_i)
\right>,
\end{equation}
where the term proportional to $\theta_0$ is a source resulting from the
random force, and the term with the minus sign is a sink resulting from the
Langevin-type damping forces.  In global equilibrium, the source and sink
terms balance one another and $\Gamma_{eqm} = 0$.
The {\em heat current} ${\bf q}$, given explicitly in equations
(\ref{microenergy}-\ref{energyflow}) of Appendix \ref{appendix1},
contains the
standard kinetic and collisional transfer contributions due to conservative
forces, as well as dissipative contributions analagous to ${\bf \Pi}_{\rm D}$
in (\ref{eqn:PT_CONTRIBS}), and ${\bf q}$ vanishes in global equilibrium.

If $\Gamma$ would be set equal to zero, equation (\ref{eqn:E_FLOW}) would
have the generic form of the energy balance equation in ordinary
hydrodynamics, where the heat current would contain a term proportional to the
temperature gradient.  As will become apparent in later sections, this is not
the case in the DPD system. Although (\ref{eqn:E_FLOW}) looks like a macroscopic 
equation for the energy balance in the presence of sources and sinks, it looses 
its physical significance after a relaxation time $t_0$, in which $e({\bf r},t) 
\rightarrow \frac{d}{2} \theta_0 n({\bf r},t)$, and (\ref{eqn:E_FLOW}) reduces to 
the continuity equation. This will be discussed in Section \ref{section6}, below 
equation (\ref{eqn:A_DEFN}).

One may also derive a balance equation for the free energy density, which would be 
a local version of the $H$-theorem of Section \ref{section3} or of the 
corresponding one of Section \ref{section5}
for the Fokker-Planck-Boltzmann equation. It would enable one to identify the 
irreversible entropy production. A similar balance equation for the entropy 
density in a dilute gas
can be derived from the Boltzmann equation \onlinecite{Groot}.

\section{Fokker-Planck-Boltzmann equation}
\label{section5}

In this section we derive an approximate kinetic equation, referred to as
the Fokker-Planck-Boltzmann (FPB) equation, for the single particle
distribution function $f({\bf x},t)$, which is based on the molecular
chaos assumption and has a collision term which is quadratic in $f({\bf x},t)$.
Moreover, from here on the {\em conservative forces} will be neglected,
which corresponds to the strong damping limit ($\gamma$ large).

This section is organised as follows.  We start by deriving the first
equation of the BBGKY hierarchy, which relates $\partial_t f$ to the
pair function $f^{(2)} ({\bf x},{\bf x}',t)$.  Then the
{\em molecular chaos} assumption:

\begin{equation} \label{eqn:MOLECULAR_CHAOS}
f^{(2)}({\bf x},{\bf x}',t) \simeq f({\bf x},t) f({\bf x}',t),
\end{equation}
yields a closed equation, the FPB equation, which again satisfies an
$H$-theorem.  Next we analyse the {\em local equilibrium} solution of the
kinetic equation, which provides the conceptual basis for the existence
of hydrodynamic equations and transport coefficients, as well as the
justification for solving this kinetic equation for finding the ``normal 
solution'' by means of the
Chapman-Enskog method.

The first equation of the BBGKY hierarchy can be derived directly by
applying equation (\ref{eqn:GRCE}) to the $\mu$-space density:

\begin{equation} \label{eqn:MU_DENSITY}
\hat{f}({\bf x}) = \sum_i \delta( {\bf x}-{\bf x}_i).
\end{equation}
Its average yields $f({\bf x},t)$ on account of (\ref{eqn:F_DEFN}).
The resulting equation of motion is :

\begin{equation}
\partial_t f = \int d{\bf \Gamma} P({\bf \Gamma},t)
\left\{ \sum_{i} \frac{\partial \hat{f}} {\partial {\bf r}_i} \cdot  {\bf v}_i 
+\sum_{i,j \neq i} \gamma w(R_{ij}) \left\{ -
{\bf v}_{ij} \frac{\partial}{\partial {\bf v}_i}
+\frac{\theta_0}{m}  \frac{\partial}{\partial {\bf v}_i}
\left( \frac{\partial}{\partial {\bf v}_i} -
\frac{\partial}{\partial {\bf v}_j} \right) \right\} \hat{f} :
{\bf \hat{R}}_{ij} {\bf \hat{R}}_{ij} \right\} ,
\end{equation}
where the $(:)$ contraction of tensors is defined by
${\sf A}:{\sf B} = \sum_{\alpha \beta} A_{\alpha \beta} B_{\beta \alpha}$,
with $\alpha$, $\beta$ denoting Cartesian components of vectors or tensors.
The equation can be further simplified to:

\begin{eqnarray}
\partial_t f &=& -\mbox{\boldmath $\nabla$} \cdot \left< \sum_i
{\bf v}_{i} \delta({\bf x}-{\bf x}_i) \right> + 
\gamma \frac{\partial}{\partial {\bf v}}\cdot  \left< \sum_{i,j \neq i} 
\delta({\bf x}-{\bf x}_i) w(R_{ij}) {\bf \hat{R}}_{ij} \left\{
{\bf \hat{R}}_{ij}\cdot {\bf v}_{ij}  \right\} \right> \nonumber \\
&+& \frac{\gamma \theta_0}{m}
\frac{\partial^{2}}{\partial {\bf v} \partial {\bf v}} :
\left< \sum_{i,j \neq i} w(R_{ij}) {\bf \hat{R}}_{ij} {\bf \hat{R}}_{ij}
\delta ({\bf x}-{\bf x}_i) \right>,
\end{eqnarray}
where $\mbox{\boldmath $\nabla$} = \partial/\partial {\bf r}$.
Performing the integrals over all variables except ${\bf x}_i$ and ${\bf x}_j$
leads to:
\begin{equation}
\partial_t f + {\bf v}\cdot \mbox{\boldmath $\nabla$} f= \gamma \int d{\bf v}' 
\int d{\bf R} \; 
{\bf \hat{R}} {\bf \hat{R}} w({\bf R}) : \left\{
\frac{\partial}{\partial {\bf v}} \left( {\bf v}-{\bf v}' \right) +
\frac{\theta_0}{m} \frac{\partial^{2}}{\partial {\bf v} \partial {\bf v}}
\right\} f^{(2)} ({\bf v},{\bf r},{\bf v}',{\bf r}-{\bf R},t).
\label{BBGKY1}
\end{equation}
This is the first equation of the BBGKY hierarchy with
the Fokker-Planck equation (\ref{eqn:FP}) taking the place of the Liouville
equation as the evolution equation. Under the molecular chaos approximation
(\ref{eqn:MOLECULAR_CHAOS}) we have the following closed equation
for the one particle distribution function:

\begin{equation} \label{eqn:BBGKY}
\partial_t f + {\bf v}.\mbox{\boldmath $\nabla$} f = I(f) \equiv 
\gamma \int d{\bf v}' \int d{\bf R} \; 
{\bf \hat{R}} {\bf \hat{R}} w(R) f({\bf v}', {\bf r}-{\bf R}, t) :
\left\{ \frac{\partial}{\partial {\bf v}} \left( {\bf v}-{\bf v}'
\right) + \frac{\theta_0}{m}
\frac{\partial^{2}}{\partial {\bf v} \partial {\bf v}} \right\}
f ({\bf v},{\bf r},t).
\end{equation}

The molecular chaos approximation is a mean-field approximation, which
neglects dynamical correlations resulting from correlated multiple
collisions taking place inside an action sphere.  As we have set all
conservative forces equal to zero, the molecular chaos assumption is
exact in the global equilibrium state.  Indeed, simulation results show that
this is in fact an excellent approximation in the small-time step limit (as 
shown in Figure 2).

\begin{figure}[h]
\epsfxsize 12cm
\epsfysize 10cm
\epsfbox{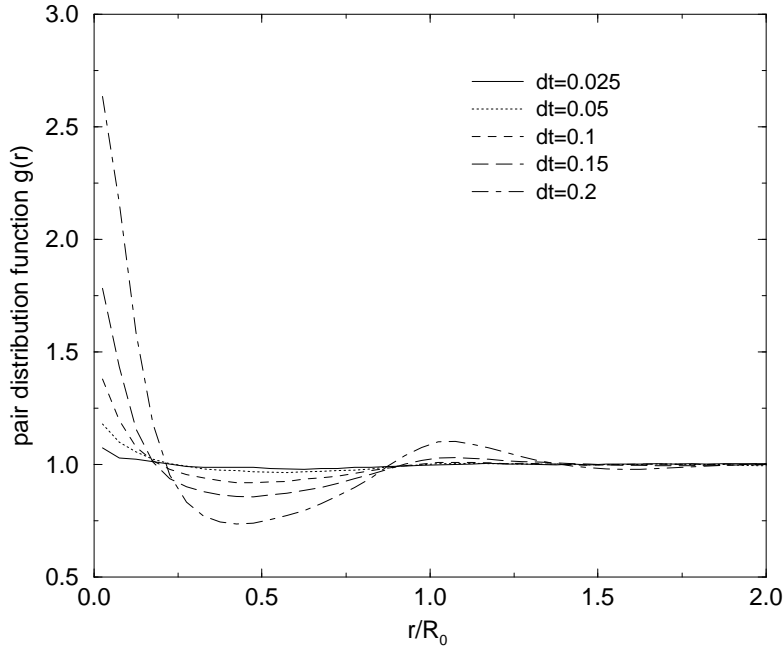}
\caption{\label{fig2} Two particle distribution function, where the separation is 
measured in units $R_0$. The system parameters in the simulation were taken:
$N=2000$ particles, friction constant $\gamma=1$, particle density $n=0.2$, action 
sphere radius $R_0=4$, and $w_{\rm D}(r)=2(1-r/R_0)$.}
\end{figure}

It can be shown in a similar fashion to section \ref{section3} that the
functional,

\begin{equation} \label{eqn:LITTLE_FUNCTIONAL}
{\cal F} = \int d{\bf x} \left\{ \frac{1}{2} m {\bf v}^2 + \theta_0
\ln f({\bf x},t) \right\} f({\bf x},t),
\end{equation}
satisfies an $H$-theorem ($\partial_t {\cal F} \leq 0$) where the equality
only holds if $f({\bf x},t)$ is given by the equilibrium form
$n_0\varphi_0({\bf v})$ of (\ref{eqn:F_EQM}), which establishes the existence
of a unique global equilibrium state.

The next problem is to solve the nonlinear FPB equation, and to analyse
the approach to equilibrium using the {\em Chapman-Enskog} (CE) method.
According to this method, one can distinguish two stages in the evolution
of the single particle distribution function $f({\bf x},t)$: a rapid kinetic
stage and a slow hydrodynamic stage \onlinecite{COHEN}.

In the {\em kinetic} stage, $f({\bf x},t)$ decays within a characteristic
kinetic time $t_0$ to the so-called {\em normal} solution $f({\bf v} \vert
a({\bf r},t))$ which depends on space and time only through the first few
moments $a({\bf r},t) = \int d{\bf v} a({\bf v}) f({\bf x},t)$, the
conserved densities, where $a({\bf v}) = \{1,{\bf v}\cdots\}$ are the
collisional invariants.  In fluid systems the time $t_0$ is the mean
free time, whereas in a DPD system, $t_0$ is estimated from (\ref{eqn:BBGKY})
as $t_0 \sim 1/(\gamma n R_0^d)$.

In the subsequent {\em hydrodynamic} stage, $f$ depends only on space and
time through its dependence on the conserved densities.  In this stage,
the solution $f({\bf v} \vert a({\bf r},t))$ of the FPB equation can be
determined perturbatively, $f=f_0+\mu f_1+\cdots$, as an expansion in powers
of a small parameter, $\mu \sim l_0 \mbox{\boldmath $\nabla$}$, which measures the 
variation
of the macroscopic parameters over a characteristic kinetic length
scale, $l_0 \simeq t_0 \bar{v} = (1/\gamma) \sqrt{\theta_0/m}$, where
$\bar{v} = \sqrt{\theta_0/m}$ is a typical mean velocity.  Therefore the
$\mu$-expansion is essentially an expansion in the small parameter
$1/\gamma$ (c.f. solution to Kramers' equation in \onlinecite{VK}).

In the remaining part of this section, we focus on determining the lowest
order solution $f_0$ of (\ref{eqn:BBGKY}), which is the {\em local equilibrium}
distribution.  We first observe that the lhs of (\ref{eqn:BBGKY}) is
of ${\cal O}(\mu)$, as $\partial_t f$ is proportional to $\partial_t a
\sim {\cal O}(\mu)$, and similarly for the gradient term.  The rhs of
(\ref{eqn:BBGKY}) is of ${\cal O}(1)$.  This requires that, to the dominant
order in $\mu$, $f$ should satisfy $I(f_0) = 0+{\cal O}(\mu)$.  To
determine the solution $f_0$, we delocalise the collision operator
$I(f_0)$ by replacing $f_0({\bf v}', {\bf r}-{\bf R},t)$ on the rhs of
(\ref{eqn:BBGKY}) by $f_0({\bf v}', {\bf r},t)+{\cal O}(\mu)$.  If we
denote the {\em delocalised} collision operator by $I_0$, then $f_0$
is the solution of:

\begin{equation} \label{eqn:f0}
I_0(f_0) = 0 \mbox{.}
\end{equation}

Guided by the $H$-theorem, we assume the standard form for the local
equilibrium distribution:

\begin{equation} \label{eqn:LOCAL_EQM}
f_0({\bf v} \vert a) = n \left( \frac{m}{2 \pi \theta} \right)^{d/2}
\exp \left[ -\frac{m({\bf v}-{\bf u})^{2}}{2\theta} \right],
\end{equation}
where $n$, $\theta$ and ${\bf u}$ are arbitrary functions of ${\bf r}$ and
$t$.  Substitution of (\ref{eqn:LOCAL_EQM}) into (\ref{eqn:f0}) shows,
however, that the above $f_o$ is {\em only} a solution if

\begin{equation} \label{eqn:TEMP_DEFN}
\theta = \theta_0 \equiv \frac{m \sigma^2}{2 \gamma} \mbox{ , }
\end{equation}
where $\theta_0$ is the constant model parameter introduced below equation
(\ref{eqn:FUNCTIONAL}), which equals the global equilibrium temperature.
The parameters $n({\bf r},t)$ and ${\bf u}({\bf r},t)$ in (\ref{eqn:LOCAL_EQM})
are chosen to be the fluid density and flow velocity.  Hence,

\begin{equation} \label{eqn:A_DEFN}
a({\bf r},t) = \int d{\bf v} a({\bf v}) f({\bf x},t) = \int d{\bf v} a({\bf v})
f_0({\bf x},t) \mbox{ , }
\end{equation}
where $a({\bf v}) = \{1,{\bf v}\}$ is a collisional invariant.

This observation (\ref{eqn:TEMP_DEFN}) has a profound consequence on the physical 
processes
occurring in the DPD system, and makes it very different from standard fluids
with energy conservation.  In fluids, there is a fast kinetic relaxation
to a local equilibrium state specified by $n({\bf r},t)$, $\theta({\bf r},t)$
and ${\bf u}({\bf r},t)$, and a subsequent slow hydrodynamic relaxation of
these fields to global equilibrium.  The DPD system distinguishes itself from
standard fluids in the sense that there is a fast relaxation on a time
scale $t_0$ to a local equilibrium state 
(\ref{eqn:LOCAL_EQM})-(\ref{eqn:TEMP_DEFN}), specified by $n({\bf r},t)$, ${\bf 
u}({\bf r},t)$
and a spatially uniform and constant temperature $\theta_0$.  The subsequent slow 
relaxation involves only the density $n({\bf r},t)$ and
flow velocity ${\bf u}({\bf r},t)$.

Consequently, a DPD system is not able to sustain a temperature gradient
on hydrodynamic time scales; there is no heat current proportional to a
temperature gradient; and there is {\em no heat conductivity}.  Thus, the DPD
system describes a {\em thermostated} or {\em isothermal} process at a
fixed temperature $\theta_0$. It may only model physical systems where the
temperature either relaxes very rapidly to an equilibrium value or where
the temperature is irrelevant (an athermal process).

It is worthwhile to explore the differences between DPD and standard fluids
somewhat further. Recalling that conservative forces have been set equal
to zero in the present situation, permits us to write the energy density,
$e({\bf r},t) = (d/2) n({\bf r},t) \theta({\bf r},t)$ in terms of a 
{\em kinetic temperature} $\theta({\bf r},t)$.  Clearly $n({\bf r},t)$
is a slowly changing variable, but what is the behaviour of
$\theta({\bf r},t)$ ?

To answer this question, it is sufficient to consider only {\em small}
deviations from global equilibrium, $\delta f = f-n_0 \varphi_0({\bf v})$,
and to linearize the FPB equation around $n_0 \varphi_0({\bf v})$.  From that
equation we shall derive how $\delta \theta({\bf r},t) = \theta({\bf r},t)
-\theta_0$ decays to zero.
Linearisation of (\ref{eqn:BBGKY}) yields, after some algebra:

\begin{equation} \label{eqn:LINEARIZED}
\partial_t \delta f + {\bf v} \cdot \mbox{\boldmath $\nabla$} \delta f  =
\omega_0 \frac{\partial}{\partial {\bf v}} \cdot  \left( {\bf 
v}+\frac{\theta_0}{m}
\frac{\partial}{\partial {\bf v}} \right) \delta f + \frac{mn_0^2\gamma}
{\theta_0} \int d{\bf R} w(R) {\bf \hat{R}}{\bf \hat{R}}:{\bf u}
({\bf r}-{\bf R}) {\bf v} \varphi_0({\bf v}) \mbox{ , }
\end{equation}
where we have used the relation
\begin{equation} \label{eqn:U0_DEFN}
n_0 {\bf u}({\bf r},t) = \int d{\bf v} \delta f({\bf x},t) {\bf v} \mbox{ , }
\end{equation}
and introduced the coefficient, $\omega_0=1/t_0$
\begin{equation} \label{eqn:OMEGA}
\omega_0 = \frac{\gamma n_0}{d} \int d{\bf R} w(R) \equiv \frac{\gamma n_0}{d}[w] 
\mbox {.}
\end{equation}
The equation with its non-local integral operator on the rhs of
(\ref{eqn:LINEARIZED}) would be the starting point for studying
generalised hydrodynamics with wave-number dependent transport coefficients.
Here, however, we shall only consider the decay of

\begin{equation} \label{eqn:THETA_DECAY}
\delta \theta({\bf r},t) = \frac{1}{n_0} \int d{\bf v}
\left( \frac{m {\bf v}^2}{d} - \theta_0 \right) \delta f \mbox{ .}
\end{equation}
Then the rate of change of $\delta f$ can be calculated from
(\ref{eqn:LINEARIZED}), and yields:

\begin{equation} \label{eqn:HEAT_MODE}
\partial_t \delta \theta + \mbox{\boldmath $\nabla$} \cdot \frac{1}{n_0}
\int d{\bf v} \left( \frac{m {\bf v}^2}{d} - \theta_0 \right) {\bf v}
\delta f = -2 \omega_0 \delta \theta \mbox{. } 
\end{equation}
The second term on the lhs of (\ref{eqn:HEAT_MODE}) is typically an
${\cal O}(\mu)$ correction to the dominant decay terms.  So
(\ref{eqn:HEAT_MODE}) is a simple relaxation equation which shows
explicitly that the kinetic temperature $\theta({\bf r},t)$ decays within
the kinetic stage to the global temperature $\theta_0$ with a relaxation
time $t_{\theta} = (2 \omega_0)^{-1} = \frac{1}{2} t_0$.

The conclusion is that the energy density in the hydrodynamic stage,
given by $e({\bf r},t) = (d/2) \theta_0 n({\bf r},t)$, is still a slow
but {\em not an independent} variable.  It is strictly proportional to the
density.  Moreover, we can conclude that the free-energy-type
functions (\ref{eqn:FUNCTIONAL}) and (\ref{eqn:LITTLE_FUNCTIONAL})
represent the actual {\em free energy} of the DPD system in the
hydrodynamic stage.

Similarly, we can determine the {\em local equilibrium} part of the pressure
tensor (\ref{eqn:PT_CONTRIBS}) in absence of conservative forces, by replacing
$f$ in ${\bf \Pi}_{\rm K}$ by its local equilibrium form $f_0$ and $f^{(2)}$
in ${\bf \Pi}_{\rm D}$ by $f_0 f_0$ according to the stosszahlansatz
\onlinecite{CC}.
To zeroth order in $\mu$, the dissipative part ${\bf \Pi}_{\rm D}$
vanishes, and the local equilibrium pressure is given by

\begin{equation} \label{eqn:LOCAL_EQM_P}
{\bf \Pi}_0 = \int d{\bf v} \; m {\bf VV} f_0 = n({\bf r},t) \theta_0
\identity.
\end{equation}
These results will be needed in the next section to solve the FPB equation
to linear order in $\mu$.

\section{Hydrodynamic Stage}
\label{section6}
\subsection{Chapman-Enskog Method}

In Section \ref{section4} we have derived the Fokker-Planck-Boltzmann equation for 
the DPD system and described the Chapman-Enskog method for obtaining its solution 
$f({\bf v}|n, {\bf u})$ in the hydrodynamical stage. The method requires that the 
rhs and lhs of the FPB-equation (\ref{eqn:BBGKY}) are expanded in powers of $\mu 
\sim l_0 \mbox{\boldmath $\nabla$}$, using the expansion

\begin{equation}
f({\bf v}|n, {\bf u})=f_0+\mu f_1+...
\label{expansion}
\end{equation}
Every $\mbox{\boldmath $\nabla$}$ is replaced by $\mu \mbox{\boldmath $\nabla$}$ 
and the derivative $\partial_t f_0$ is eliminated using the macroscopic 
conservation laws.
The lowest order solution, which is the local equilibrium distribution,
\begin{equation}
f_0= n({\bf r},t) \left( \frac{m}{2 \pi \theta_0} \right)^{d/2}
\exp \left[-\frac{m}{2 \theta_0}
({\bf v}-{\bf u}({\bf r},t))^{2} \right] \mbox{ , }
\label{localequi}
\end{equation}
has been determined in the previous section.

To obtain $f_1$ we expand the FPB equation in powers of $\mu$, yielding
\begin{equation}
\frac{\partial f_0}{\partial t}+ \mu {\bf v}\cdot \mbox{\boldmath $\nabla$} 
f_0=I(f_0)+ \mu (dI/df)_{f_o} f_1+...
\label{BM}
\end{equation}
We start with the rhs of (\ref{BM}), which has been calculated exactly
to ${\cal O}(\mu)$-terms included. One finds after some algebra that
$I(f_0)={\cal O}(\mu^2)$. In the previous section it has only been
verified that $I(f_0)={\cal O}(\mu)$. The latter result is {\em not}
sufficient, whereas the former is sufficient for our present
purpose. In the remaining terms on the rhs we replace the collision
operator $I$ by its delocalised form $I_0$, as defined below
(\ref{eqn:f0}). The rhs of (\ref{BM}) then becomes: 
\begin{equation} 
(dI/df)_{f_0}f_1 = \omega_0
\frac{\partial}{\partial {\bf V}}\cdot 
\left( {\bf V}+\frac{\theta_0}{m} \frac{\partial}{\partial {\bf V}} \right)f_1 
\mbox{ ,}
\label{enskog1M}
\end{equation}
with $\omega_0$ defined in (\ref{eqn:OMEGA}).

To calculate the lhs of (\ref{BM}) to ${\cal O}(\mu)$ we need the rate of change 
of $n$ and ${\bf u}$ to lowest order in $\mu$, which may be calculated from the 
conservation equations (\ref{eqn:MACRO_CONS}) with ${\bf \Pi}$ replaced by its 
local equilibrium part ${\bf \Pi}_0=n \theta_0 {\identity}$, calculated in 
(\ref{eqn:LOCAL_EQM_P}), i.e.
\begin{eqnarray} 
\partial_t n &=&-\mbox{\boldmath $\nabla$} \cdot  (n {\bf u})
\label{cont}\\
\partial_t {\bf u} &=& - {\bf u}\cdot  \mbox{\boldmath $\nabla$} {\bf u} - 
\frac{\theta_0}{\rho} \mbox{\boldmath $\nabla$} n \mbox{. }
\nonumber
\end{eqnarray}
They yield in combination with (\ref{localequi})
\begin{equation}
\frac{\partial f_0}{\partial t}+ {\bf v}\cdot   \mbox{\boldmath $\nabla$} f_0=f_0 
[{\bf \sf J}:{\bf \sf D}+{\cal J} \mbox{\boldmath $\nabla$} \cdot  {\bf u}]
\label{split}
\end{equation}
with
\begin{eqnarray}
{\bf \sf J}_{\alpha \beta}(V) &=& \frac{m}{\theta_0} \left\{ V_{\alpha} V_{\beta} 
- \frac{1}{d} \delta_{\alpha \beta} V^2 \right\} \label{defBM} \\
{\bf \sf D}_{\alpha \beta}(V) &=& \frac{1}{2} \left\{ \mbox{\boldmath 
$\nabla$}_{\alpha} u_{\beta} + \mbox{\boldmath $\nabla$}_{\beta} u_{\alpha} 
-\frac{2}{d}
 \delta_{\alpha \beta} \mbox{\boldmath $\nabla$} \cdot  {\bf u} \right\} \mbox{ , 
} \nonumber
\end{eqnarray}
and 
\begin{equation}
{\cal J}(V)= \frac{m V^2}{d \theta_0} -1 \mbox{. }
\label{def2BM}
\end{equation}
Note that the density and temperature gradients are absent on the rhs of 
(\ref{split}), in contrast to the traditional Chapman-Enskog result
\cite{CC}.

For convenience of notation we introduce the Fokker-Planck operator ${\cal L}$ and 
its adjoint ${\cal L}^{+}$, defined as
\begin{eqnarray}
{\cal L} &=& \frac{\partial}{\partial {\bf V}} \cdot  [{\bf V}+ 
\frac{\theta_0}{m}\frac{\partial}{\partial {\bf V}}]\label{operators}\\
{\cal L}^{+} &=&  [- {\bf V}+ \frac{\theta_0}{m} \frac{\partial}{\partial {\bf 
V}}]\cdot \frac{\partial}{\partial {\bf V}},  \nonumber
\end{eqnarray}
to write the final equation for $f_1$ as:
\begin{equation}
\omega_0 {\cal L}f_1=f_0 [{\bf \sf J}:{\bf \sf D}+{\cal J} \mbox{\boldmath 
$\nabla$} \cdot  {\bf u}] \mbox{. }
\label{finaleq}
\end{equation}
It is a second order PDE with an inhomogeneity on the rhs. We first construct a 
special solution by recalling that the Fokker-Planck operator ${\cal L}$ can be 
mapped onto the Schr\"{o}dinger equation for an isotropic $d$-dimensional harmonic 
oscillator \onlinecite{VK}. Its eigenfunctions are the tensor Hermite polynomials, 
usually called Sonine polynomials in a kinetic theory context, and the microscopic 
fluxes ${\bf \sf J}$ and ${\cal J}$ are among them, i.e.
\begin{eqnarray}
{\cal L} f_0 {\bf \sf J} = - 2 f_0 {\bf \sf J} \qquad \mbox{ ; } {\cal L}^{+}{\bf 
\sf J} = - 2 {\bf \sf J} \label{eigenfuncties}\\
{\cal L} f_0 {\cal J}= - 2 f_0 {\cal  J} \qquad \mbox{ ; } {\cal L}^{+} {\cal J} = 
- 2 {\cal J} \mbox{.} \nonumber
\end{eqnarray}
This can easily be verified. Combination of (\ref{finaleq}) and 
(\ref{eigenfuncties}) yields the special solution
\begin{equation}
f_1 = -\frac{1}{2 \omega_0} f_0 [{\bf \sf J}:{\bf \sf D}+{\cal J} \mbox{\boldmath 
$\nabla$}\cdot  {\bf u}] \mbox{.}
\label{specsol}
\end{equation}
The general solution is obtained by adding an arbitrary linear combination of 
collisional invariants $a({\bf V})=\left\{ 1,{\bf V} \right\}$, which are the 
solutions to the homogeneous equation ${\cal L} f_0 a({\bf V})=0$. However, the 
constraint (\ref{eqn:A_DEFN}) suppresses these terms and $f_1$ is the desired 
solution of the FPB equation to linear order in $\mu$.

\subsection{Navier-Stokes Equation}
The only slow macroscopic fields are the density $n({\bf r},t)$ and the flow 
velocity ${\bf u}({\bf r},t)$, leading to the continuity equation and 
Navier-Stokes equation. The energy density in the hydrodynamic stage, $e({\bf 
r},t)=(d/2) n({\bf r},t) \theta_0$, is not an independent variable. The energy 
balance equation derived in the Appendix \ref{appendix1} is only relevant in the 
kinetic stage, but has no physical significance in the hydrodynamical stage.

The results for $f_0$ and $f_1$ are sufficient to obtain the hydrodynamic 
equations to Navier-Stokes order and to obtain explicit expressions for the 
transport coefficients. The ${\cal O}(\mu)$-correction $f_1$ contains only 
gradients of the flow field, $\mbox{\boldmath $\nabla$} {\bf u}$, but {\it {no}} 
gradients of the temperature. Therefore, there will be no heat current and 
vanishing heat conductivity.
The ${\cal O}(\mu)$-terms in the pressure tensor ${\bf \Pi}= {\bf \Pi}_0+\mu {\bf 
\Pi}_1+...$ will be proportional to $\mbox{\boldmath $\nabla$} {\bf u}$ and we 
define the viscosities as the coefficients of proportionality through the 
constitutive relation
\begin{equation}
{\bf \Pi}_1=-2 \eta {\bf \sf D}-\zeta \mbox{\boldmath $\nabla$} \cdot  {\bf u} 
{\identity},
\label{defvis}
\end{equation}
where $\eta$ and $\xi$ are respectively shear and bulk viscosity.

Combining (\ref{defvis}), (\ref{eqn:LOCAL_EQM_P}) and
(\ref{eqn:MACRO_CONS}) then yields the Navier-Stokes equation for the DPD system 
\begin{equation}
\partial_t  (\rho {\bf u}) +\mbox{\boldmath $\nabla$} \cdot  (\rho {\bf u}{\bf 
u})=-\theta_0 \mbox{\boldmath $\nabla$} n + \mbox{\boldmath $\nabla$} \cdot  (2 
\eta {\bf \sf D}+ \zeta \mbox{\boldmath $\nabla$} \cdot  {\bf u} \identity).
\label{ns}
\end{equation}
The explicit expressions will be obtained in the next section.

\section{Transport Coefficients}
\label{section7}
\subsection{Kinematic viscosities $\eta_{\rm K}$ and $\zeta_{\rm K}$} 
There are two contributions to the pressure tensor: the {\em kinetic} part ${\bf 
\Pi}_{\rm K}$ and the {\em dissipative} part ${\bf \Pi}_{\rm D}$, defined in 
(\ref{eqn:PT_CONTRIBS}) and two corresponding viscosities. The kinetic part 
depends only on $f=f_0+\mu f_1$, which are given in (\ref{localequi}) and 
(\ref{specsol}).  Then ${\bf \Pi}_{\rm K}$ becomes 
\begin{equation}
{\bf \Pi}_{\rm K}=n \theta_0 {\identity}+\mu {\bf \Pi}_{{\rm K},1}+...
\label{kinpart}
\end{equation}
where $\mu$ is a formal expansion parameter that will be set equal to unity at the 
end of the calculations and

\begin{equation}
{\bf \Pi}_{{\rm K},1}=\int d{\bf v} m {\bf V} {\bf V} f_1= \theta_0 \int d{\bf v} 
\left\{ {\bf \sf J}({\bf V})+{\cal J}({\bf V}) {\identity} \right\} f_1  \mbox{.}
\label{kinpart1}
\end{equation}
Here the dyadic $m {\bf V}{\bf V}$ has been split up into a traceless tensor, 
$\theta_0 {\bf \sf J}$, and a term $\theta_0 {\cal J}{\identity}$, proportional to 
the unit tensor and we have used the relation $\int d{\bf v} f_1 =0$ (see 
(\ref{eqn:A_DEFN})).
Inserting the explicit solution (\ref{specsol}) into (\ref{kinpart1}) allows us to 
write (\ref{kinpart}) in the form   
\begin{equation}
{\bf \Pi}_{{\rm K},1} = -\frac{n \theta_0 }{2\omega_0} <{\bf \sf J}|{\bf \sf \bf 
\sf J}>: {\bf \sf D} - \frac{n \theta_0 }{2\omega_0} <{\cal J}|{\cal J}>  
\mbox{\boldmath $\nabla$} \cdot  {\bf u} \identity \mbox { , }
\label{pressuretensor}
\end{equation}
where we have used the relation $f_0=n \varphi_0(V)$ (see (\ref{eqn:F_EQM})) and 
introduced the inner product
\begin{equation}
<A|B> =  \int d{\bf v} \varphi_0(V) A(V) B(V) \mbox{.}
\label{defbracket}
\end{equation}
Moreover a crossproduct of a traceless tensor and a scalar vanishes , i.e. $<{\bf 
\sf J}|{\cal J}>=0$.
The product $<{\cal J}|{\cal J}>$ involves a simple Gaussian integral and yields
\begin{equation}
<{\cal J}|{\cal J}>=\frac{2}{d} \mbox{.}
\label{inprod}
\end{equation}
The fourth rank tensor $<{\sf J}|{\sf J}>$ is isotropic, traceless and symmetric, 
which implies the general form
\begin{eqnarray}
<{\bf \sf J}_{\alpha \beta}|{\bf \sf J}_{\delta \gamma}> &=& \left( 
\frac{m}{\theta_0} \right)^2 \int d{\bf v} \varphi_0(V) \left( 
V_{\alpha}V_{\beta}-\frac{1}{d} \delta_{\alpha \beta} V^2 \right) V_{\gamma} 
V_{\delta} \label{stream} \\
&=& C \left[ \delta_{\alpha \gamma} \delta_{\beta \delta}+\delta_{\alpha \delta} 
\delta_{\beta \gamma}-\frac{2}{d}\delta_{\alpha \beta} \delta_{\gamma \delta} 
\right] \mbox{.} \nonumber
\end{eqnarray}
By taking double contractions and evaluating Gaussian integrals the constant $C$ 
comes out to be equal to $1$ and the $(:)$ product in 
(\ref{pressuretensor}) yields
\begin{equation}
<{\bf \sf J}|{\bf \sf J}> : {\bf \sf D}= 2 {\bf \sf D} \mbox{.}
\label{eigenvalue}
\end{equation}
Combination of (\ref{pressuretensor}), (\ref{inprod}) and (\ref{eigenvalue}) 
finally yields
\begin{equation}
{\bf \Pi}_{{\rm K},1}= -\frac{n \theta_0}{\omega_0} {\bf \sf D}-\frac{n 
\theta_0}{d \omega_0} \mbox{\boldmath $\nabla$} \cdot  {\bf u} {\identity} 
\mbox{.}
\label{result}
\end{equation}
Comparison with the constitutive relation (\ref{defvis}) enables us to identify 
the coefficients as the {\it kinetic parts} of the viscosities
\begin{equation}
\eta_{\rm K} = \frac{n \theta_0}{2 \omega_0} =\frac{d \theta_0}{2 [w]} 
\qquad
\zeta_{\rm K} = \frac{n \theta_0}{d \omega_0}= \frac{\theta_0}{\gamma [w]}\mbox{ , 
} 
\label{transport}
\end{equation}
where the definition (\ref{eqn:OMEGA}) of $\omega_0$ has been used. We note that 
the kinetic part is inversely proportional to $\gamma$ and has been for the first 
time explicitly calculated. 

\subsection{Dissipative viscosities $\eta_{\rm D}$ and $\zeta_{\rm D}$}
This section deals with the dissipative part ${\bf \Pi}$ of the pressure tensor in 
(\ref{eqn:PT_CONTRIBS}), which depends on the pair distribution function 
$f^{(2)}$. This function has a local equilibrium part $f_0^{(2)}$ and a part $\mu 
f_1^{(2)}$, linear in the gradients. We start with the first part.   

In order to make a direct comparison with the work of Espa\~{n}ol 
\onlinecite{PE:HD} we retain the conservative forces, for the time being.
Then, the local equilibrium pair function has the form
\begin{equation}
f_0^{(2)}({\bf x},{\bf x}')=f_0({\bf x}) f_0({\bf x}') g_0(\vert {\bf x}-{\bf x}' 
\vert) ,
\label{loc_eq_pair}
\end{equation}
where $g_0(R)$ is the spatial correlation function in local equilibrium. Only at 
the end of the calculation we will set the conservative forces equal to zero, so 
that $g_0(R)=1$.

Substitution of (\ref{loc_eq_pair}) into (\ref{eqn:PT_CONTRIBS}) yields then

\begin{eqnarray}
{\bf \Pi}_{\rm D} &=&- {\textstyle{\frac{1}{2}}} \gamma m \int d{\bf R} g_0(R) 
w(R) {\bf \hat{R}}{\bf \hat{R}}
\left( {\bf R} \cdot  [{\bf u}({\bf r})-{\bf u}({\bf r}-{\bf R})] \right) n({\bf 
r}) n({\bf r}-{\bf R}) \label{dispart}\\
&\simeq& - {\textstyle{\frac{1}{2}}} \gamma m n^2 \int d{\bf R} R^2 g_0(R) w(R) 
{\bf \hat{R}} {\bf \hat{R}} {\bf \hat{R}} {\bf \hat{R}}: \mbox{\boldmath $\nabla$} 
{\bf u} \mbox{ , } \nonumber
\end{eqnarray}
where $[...]$ has been expanded to linear order in the gradients. Calculation of 
the completely symmetric isotropic fourth rank tensor proceeds as in 
(\ref{stream}) with the result
\begin{equation}
\int d{\bf R} R^2 g_0(R) w(R) \hat{R}_{\alpha} \hat{R}_{\beta} \hat{R}_{\gamma} 
\hat{R}_{\delta}= \frac{[R^2 w g_0]}{d(d+2)}[\delta_{\alpha \beta} \delta_{\gamma 
\delta}+
\delta_{\alpha \delta} \delta_{\beta \gamma }+\delta_{\alpha \gamma} \delta_{\beta 
\delta}] \mbox{ , }
\label{contrac}
\end{equation}
where 
\begin{equation}
[R^2 w g_0] \equiv \int d{\bf R} R^{2} g_0(R) w(R) \mbox{.}
\label{moment}
\end{equation}
We note that the definition of ${\bf \Pi}_{\rm D}$ in (\ref{eqn:PT_CONTRIBS}) 
contains $\overline{f}^{(2)}$ rather than $f^{(2)}({\bf v},{\bf r},{\bf v'},{\bf 
r}-{\bf R},t)$. One easily verifies that the spatial averaging, denoted by the 
overline, makes no difference to linear order in the gradients. The final result 
for the dissipative part then becomes
\begin{equation}
{\bf \Pi}_{\rm D}= - \frac{m \gamma n^2 [R^2 w g_0]}{d(d+2)} {\bf \sf D}- \frac{m 
\gamma n^{2} [R^2 w g_0]}{2d^{2}} \mbox{\boldmath $\nabla$}\cdot  {\bf u} 
{\identity}.
\label{resultdis}
\end{equation}
With the help of (\ref{defvis}) the coefficients can be identified as the 
contributions to the viscosities due to the dissipative forces, i.e.
\begin{equation}
\eta_{\rm D} = \frac{m \gamma n^2 [R^2 w g_0]}{2d(d+2)} \qquad
\zeta_{\rm D} = \frac{m \gamma n^2 [R^2 w g_0]}{2d^2}\mbox{.} 
\label{transport2}
\end{equation}
The local equilibrium contribution (\ref{resultdis}) to the dissipative pressure 
tensor turns out to be the dominant contribution to the viscosity of a
DPD fluid, for large values of $n \gamma$, as illustrated in Figure 3
and confirmed by numerical simulation in \onlinecite{HK:DPD,ME1}. We also
want to point out that an ${\cal{O}}(\mu)$-contribution to the pressure tensor, 
calculated in local equilibrium as in (\ref{dispart}), is not a novelty of this 
paper, but also occurs in all systems with impulsive (hard core) interactions that 
are {\em not strictly local}. 
For instance, consider the collisional transfer contribution analogous to ${\bf 
\Pi}_{\rm C}$ for elastic hard spheres, where 
${\bf F}=-\mbox{\boldmath $\nabla$} \phi$ in (\ref{eqn:PT_CONTRIBS}) is 
ill-defined. This term is calculated in sections 16.4 and 16.5 of \onlinecite{CC}, 
where its local equilibrium contribution yields $\eta_{\rm {HS}}=\frac{3}{5} 
\zeta_{\rm {HS}}=\frac{3}{5} \varpi$ with $\varpi \sim n^2$, defined in Eq. 
(16.5.7) of \onlinecite{CC}. These contributions in real fluids are the direct 
counterparts of $\eta_{\rm D}=\frac{3}{5} \zeta_{\rm D} \sim n^2$ in DPD.

We return to the DPD system {\em without} conservative forces, where the Gibbs' 
distribution (\ref{eqn:GIBBS}) reduces to 
$\prod_{i=1}^{N} \varphi_o(v_i)$, and where the spatial correlations
are absent, i.e. $g_0(R)=1$. This equality is also required here for
consistency with the molecular chaos approximation
(\ref{eqn:MOLECULAR_CHAOS}), used in section \ref{section5} and subsequent ones.

So far, we have only considered local equilibrium contributions to ${\bf \Pi}_D$ 
in (\ref{dispart}). To obtain the complete contribution, consistent with the 
molecular chaos assumption, we substitute (\ref{eqn:MOLECULAR_CHAOS}) into 
(\ref{eqn:PT_CONTRIBS}) and use the definitions (\ref{eqn:FIELD_DEFN}). 
Surprisingly, the results (\ref{dispart}) are recovered, showing that 
(\ref{dispart})-(\ref{transport2}) give the full contribution of ${\bf \Pi}_{\rm 
D}$ to the Navier-Stokes equation, at least within the molecular chaos assumption.

To facilitate the comparison with the original predictions of 
\onlinecite{HK:DPD,ME1,VDB}, we set $g_0=1$ in (\ref{resultdis}) and introduce
\begin{equation}
<R^2>_{w}=  [R^2 w]/  [w],
\label{defweight}
\end{equation}
where $[a]$ denotes the spatial average introduced in (\ref{eqn:OMEGA}), so that 
$<R^2>_{w} \sim  R_0^2$.     

The final result for the dissipative part of the viscosities is then
\begin{eqnarray}
\eta_{\rm D} &=& \frac{\gamma m n^2 <R^2>_{w} [w]}{2d(d+2)} = \omega_0 t_w^2 n 
\theta_0/2(d+2) \label{dispartvis}\\
\zeta_{\rm D} &=& \frac{\gamma m n^2 <R^2>_{w} [w]}{2d^2} = \omega_0 t_w^2 n 
\theta_0/2d, \nonumber
\end{eqnarray}
where $\omega_0=1/t_0=\gamma n [w]/d$ is the characteristic relaxation rate 
introduced in (\ref{eqn:OMEGA}), and $t_w$, defined through $t_w^2= \break 
<R^2>_{w}\bar{v}^2$, is the average trasversal time of an action sphere with 
${\bar v}=(\theta_0/m)^{1/2}$ the thermal velocity. These results are in fact the 
theoretical predictions for the {\em total} shear and bulk viscosity of the DPD 
fluid, as obtained in \onlinecite{ME1,VDB} on the basis of the ``continuum 
approximation'' to the equations of motion of the DPD particles. In the present 
context of non-equilibrium statistical mechanics and kinetic theory, these 
contributions have been identified as the local equilibrium contributions to the 
transport coefficients in order to make the connection with Hoogerbrugge and 
Koelman's expression for the kinematic viscosity:   
\begin{equation}
\nu = \eta/\rho =\frac{\omega <R^2>_{w}}{2d(d+2) \delta t}
\label{KMHB}
\end{equation}
with their friction constant $\omega=\gamma \delta t$, proportional to $\delta t$, 
as a proper friction should. Moreover, we recall that the range function $w(R)$ in 
\onlinecite{HK:DPD} is normalised as,
\begin{equation}
n[w]= n \int d{\bf R} w(R) = 1.
\label{norm}
\end{equation}
So, the results (\ref{dispartvis}) and (\ref{KMHB}) are identical. Hoogerbrugge 
and Koelman have also shown that the viscosity found in their numerical 
simulations approaches (\ref{dispartvis}) and (\ref{KMHB}) for large $n \gamma$. 
Simulations carried out with the modified DPD algorithm show the same properties 
\onlinecite{ME1}.

We conclude this subsection by listing the full results (\ref{transport}) and 
(\ref{dispartvis}) for the shear and bulk viscosity in a DPD fluid with continuous 
time ($\delta t \rightarrow 0$):

\begin{eqnarray}
\eta=\eta_{\rm D}+\eta_{\rm K} &=&
\frac{1}{2} n \theta_0 \left\{ \frac{\omega_0 t_w^2}{d+2} + \frac{1}{\omega_0} 
\right\}      
\label{shearvis}\\
\zeta=\zeta_{\rm D}+\zeta_{\rm K} &=&  \frac{1}{d} n \theta_0 \left\{ 
\frac{\omega_0 t_w^2}{2} + \frac{1}{\omega_0} \right\} \nonumber
\end{eqnarray}
They involve the two intrinsic time scales of the DPD fluid: the characteristic 
kinetic time $t_0=1/\omega_0$ (see (\ref{eqn:OMEGA})) and the traversal time $t_w$ 
of an action sphere, as defined below (\ref{dispartvis}), which is of order 
$R_0/\bar{v}$.

In the parameter range $t_w > t_0$ the estimates $\eta_D$ and
$\zeta_D$ of \onlinecite{ME1,VDB} dominate, and in the range $t_w <
t_0$ the kinematic viscosities do, as illustrated in Figure 3. 

\subsection{Numerical simulations}

As a simple test, the shear viscosity $\eta$ of the DPD system was
measured in two dimensions using a
physical method.  A linear velocity gradient was established between
two moving plates and the force required to maintain this system was
measured once equilibrium had been attained.

By means of these simulations, we have measured the viscosity of the
DPD fluid as a function of $n \gamma$ at different temperatures
$\theta_0=m\sigma^2/2 \gamma$. Results are shown in Figure 3 for a higher 
temperature to emphasize the importance of the kinematic contribution.
At large $n \gamma$ the measured viscosity approaches the theoretical
prediction when the time step $\delta t$ is reduced. In this 
range of parameters, the viscosity is dominated by its dissipative part
(\ref{dispartvis}), corresponding to the original estimates of
Hoogerbrugge and Koelman.
At small $n \gamma$ and high temperature $\theta_0$ the viscosity is dominated by 
the kinetic viscosity.

At small $n \gamma$ there are sizeable differences 
between predicted and simulated results, which do not decrease with decreasing 
time step size. The breakdown of the theory in this range of parameters has  
a fundamental reason. Inspection of the collision term on the rhs of
(\ref{eqn:BBGKY}) or (\ref{enskog1M}) shows that with $\omega_0 \sim n \gamma$ and 
$\omega_0 \theta_0 \sim \sigma^2$ small the typical size of the collision term 
$\sim 1/t_0$ may not be large compared to the propagation terms on the rhs of 
(\ref{eqn:BBGKY}). Consequently, the Chapman-Enskog expansion will be poorly 
convergent or even divergent, because the kinetic and hydrodynamc time regimes are 
no longer well separated, or, equivalently, because the change of the macroscopic 
flow velocity over the characteristic kinetic length scale becomes large. To be 
consistent with the physical requirement of well separated time scales in this 
range of parameters, the imposed velocity gradients have to be reduced.

Remaining differences between theory and simulations may be caused 
by a breakdown of the molecular chaos assumption, which neglects the dynamical 
correlations that may have to be taken into account.

\begin{figure}[h]
\epsfxsize 12cm
\epsfysize 10cm
\epsfbox{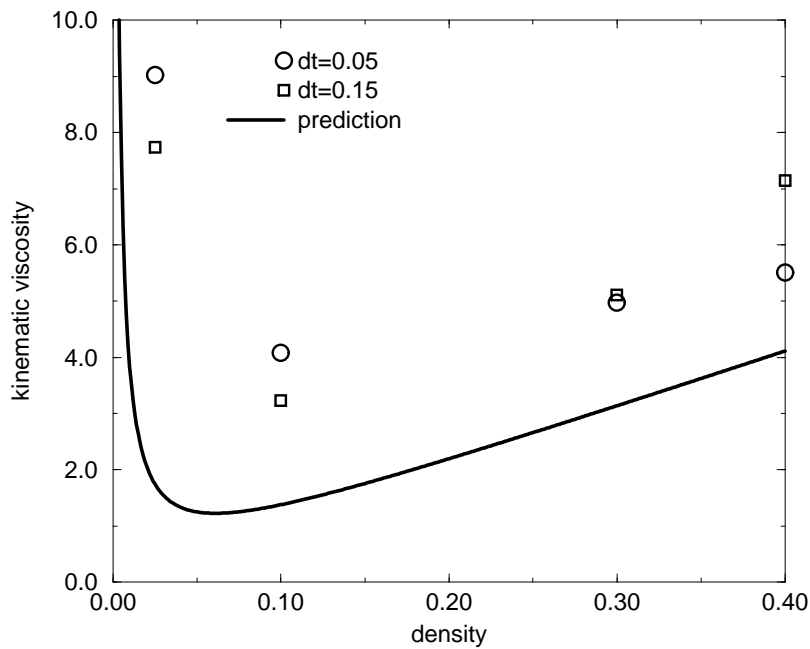}
\caption{\label{fig3} Kinematic viscosity $\nu=\eta/\rho$ against density $n$ for
$dt=0.05$ and $dt=0.15$. The system parameters in the simulations were taken :
friction constant $\gamma=1$ and strength random force $\sigma=1.5$ for densities $n=0.025$, $0.1$, $0.3$ and $0.4$.}
\end{figure}

\subsection{Self-diffusion coefficient $D$}
The coefficient of self-diffusion can be obtained by considering a DPD fluid that 
is in equilibrium, except for the probability 
distribution $f_s$ of a tagged particle, labeled as $i=1$. Following the arguments 
of Section \ref{section5} and choosing the $\mu$-space density
$\hat f_{s}({\bf x})=\delta({\bf x}-{\bf x}_1)$, instead of 
(\ref{eqn:MU_DENSITY}), one arrives at an equation similar to (\ref{BBGKY1}), with 
$f$ and $f^{(2)}$ replaced by $f_s$ and $f_s^{(2)}$ respectively, defined as
\begin{eqnarray}
f_s({\bf x},t) &=& <\delta({\bf x}-{\bf x}_1)> \label{distrifs}\\
f_s^{(2)}({\bf x},{\bf x'},t) &=& <\sum_{j \neq 1} \delta({\bf x}-{\bf x}_1) 
\delta({\bf x}-{\bf x}_j)>. \nonumber
\end{eqnarray}
The molecular chaos assumption (\ref{eqn:MOLECULAR_CHAOS}) now takes the form  
\begin{equation} 
f_s^{(2)}({\bf x},{\bf x'},t)=n \varphi_o(v')f_s({\bf x},t),
\label{molchaos2}
\end{equation}
as the fluid particles are in thermal equilibrium with the Maxwellian 
$\varphi_o(v)$ defined in (\ref{eqn:F_EQM}), and the resulting FPB equation is 
linear, i.e.
\begin{equation}
\partial_t f_s +{\bf v} \cdot \mbox{\boldmath $\nabla$} f_s= \omega_0 
\frac{\partial}{\partial {\bf v}} \cdot [{\bf v}+\frac{\theta_0}{m} 
\frac{\partial}{\partial {\bf v}}] f_s.
\label{fpe2}
\end{equation}
It is identical to the Kramers' equation (\ref{eqn:KRAMERS}) with ${\bf F}({\bf 
r})=0$.

The continuity equation takes the form
\begin{equation}
\partial_t c({\bf r},t) + \mbox{\boldmath $\nabla$} \cdot {\bf j}({\bf r},t) =0 ,  
\label{cont2}
\end{equation}
with tagged particle density and current defined as 
\begin{eqnarray}
c({\bf r},t) &=& \int d{\bf v} f_s({\bf x},t), \label{deftagcurden} \\
{\bf j}({\bf v},t) &=& \int d{\bf v} {\bf v} f_s({\bf x},t). \nonumber
\end{eqnarray}
Application of the Chapman-Enskog method to (\ref{fpe2}) yields the ``local 
equilibrium'' distribution function 
$f_{so}=c({\bf r},t) \varphi_o(v)$ and following equation for $f_{s1} \equiv 
f_s-f_{s0}$,
\begin{equation}
\varphi_o(v) {\bf v} \cdot \mbox{\boldmath $\nabla$} c = \omega_0 
\frac{\partial}{\partial {\bf v}} \cdot ({\bf v}+ \frac{\theta_0}{m} 
\frac{\partial}{\partial {\bf v}}) f_{s1} \equiv \omega_0 {\cal L} f_{s1}.
\label{firstapproxfs}
\end{equation}
As ${\bf v} \phi_o$ at the lhs is again an eigenfunction of ${\cal L}$ with 
eigenvalue $-1$, we find
\begin{equation}
f_{s1}= - \frac{1}{\omega_0} \phi_o(v) {\bf v} \cdot \mbox{\boldmath $\nabla$} c.
\label{fs1}
\end{equation}
The coefficient of {\em self-diffusion} $D$, defined through the constitutive 
equation:
\begin{equation}
{\bf j}=\int d{\bf v} {\bf v} f_1 = - D \mbox{\boldmath $\nabla$} c,
\label{defselfdif}
\end{equation}
becomes for the DPD fluid:
\begin{equation}
D= \frac{\theta_0}{\omega_0 m} =\frac{d \theta_0}{\rho \gamma [w]},
\label{selfdif}
\end{equation}
where $\rho= m n$ is the mass density of the fluid. The above result is new.
There is only a kinetic contribution and no dissipative one.  

\section{Conclusions and prospects}
\label{section8}
The main results of this paper are the derivation and solution of the
Fokker-Planck-Boltzmann (FPB) equation for the DPD fluid, providing
explicit results for the thermodynamic and transport properties in
terms of the system parameters: density $n$, friction constant
$\gamma$, temperature $\theta_0 = m \sigma^2 /2\gamma$ and range function $w(R)$ 
with range $R_0$. There are two intrinsic time scales: the kinetic relaxation time 
$t_0 \sim 1/n \gamma R_0^d$ determined by the collision term, and the traversal 
time $t_w \sim R_0/\bar{v}$ of an action sphere, where $\bar{v}=\sqrt{\theta_0/m}$ 
is the average velocity. We highlight the most important new results and future 
prospects in a number of comments.

\begin{enumerate}
\item
The DPD fluid for continuous time (step size $\delta t \rightarrow 0$), described 
by the $N$-particle Fokker-Planck equation of Espa\~{n}ol and Warren, obeys an 
$H$-theorem for the free energy ${\cal F}$. The indispensable role of the detailed 
balance condition in establishing such a theorem is demonstrated. It guarantees a 
monotonic approach of ${\cal F}$ towards a unique thermal equilibrium, described 
by the Gibbs' distribution with a temperature $\theta_0=m \sigma^2 /2 \gamma$.

\item
The local conservation laws for mass density $\rho=nm$ and momentum density are 
the essential prerequisites for the validity of the Navier-Stokes equations. The 
temperature, however, plays a very peculiar role. On the one hand the detailed 
balance condition guarantees the existence of a well-defined thermal equilibrium 
with a global equilibrium $\theta_0$, in which energy is conserved on average. On 
the other hand, the local equilibrium state depends only on $n({\bf r},t)$ and 
${\bf u}({\bf r},t)$, but not on a local equilibrium temperature $\theta({\bf 
r},t)$, which relaxes in a time $t_0$ (kinetic stage) towards its uniform 
equilibrium value $\theta_0$. In the subsequent hydrodynamic stage the DPD fluid 
is not able to sustain a temperature gradient, there is no heat conduction and all 
processes occur isothermally.

\item
In the coarse-grained mesoscopic interpretation of DPD particles
as ``lumps of fluids'',  the microscopic conservative forces between
the DPD particles are small compared to the mesoscopic friction and
random noise (large $\gamma$ limit), and have been neglected in
deriving the FPB equation. At sufficiently low temperature conservative forces can have the effect of forcing the DPD particles into crystalline configurations.

\item 
The FPB equation is derived from the first equation of the BBGKY-hierarchy for the 
distribution functions, obtained from the $N$-particle Fokker-Planck equation, 
playing the role of the Liouville equation. In addition, the molecular chaos 
assumption, $f^{(2)}({\bf x},{\bf x'})=f({\bf x})f({\bf x'})$ has been used.

\item
The Chapman-Enskog solution to the continuous time FPB equation yields two types 
of contributions to the viscosities: (i) {\em Dissipative} parts $\eta_{\rm D}$ 
and $\zeta_{\rm D}$, accounting for the collisional transfer through the nonlocal 
dissipative interactions. They are determined by the local equilibrium 
distribution. (ii) Kinetic parts $\eta_{\rm K}$ and $\zeta_{\rm K}$, coming from 
the collision operator and determined by the Chapman-Enskog solution of the FPB 
equation. If $t_w > t_0$ the dissipative viscosities are dominant; if $t_w < t_0$ 
the kinetic viscosities are dominant.
In \onlinecite{ME1,VDB} the total viscosity is estimated by $\eta_{\rm D}$ and 
$\zeta_{\rm D}$, which is correct for $t_w \gg t_0$. The simulated results for the 
kinematic viscosity are in reasonably good agreement with the predictions within 
the assessed theoretical regions of validity of the theory. We also calculated the 
coefficient of self-diffusion, which only has a kinetic part.

\item
It is also of interest to consider the Green-Kubo formulae for the
viscosities in a DPD fluid as derived in \onlinecite{PE:HD}, where the
linear $\gamma$-dependence of the viscosity in the limit of {\em large}
dissipation $\gamma$ is questioned. To make the connection we observe
that $\eta_{\rm D}$ and $\eta_{\rm C}$ of \onlinecite{PE:HD} should be
identified respectively with $\eta_{\rm D}$ and $\eta_{\rm K}$ of the present
paper. The time correlation functions of \onlinecite{PE:HD} for
$\eta_{\rm D}$ and $\eta_{\rm C}$ are formally proportional to
$\gamma^2$ and $1$ respectively. Both time integrals appearing in the
Green-Kubo formulae  extend over the characteristic kinetic time to $\sim 
1/\gamma$. Consequently, $\eta_{\rm D} \sim \gamma$ and $\eta_{\rm C} \sim 
1/\gamma$ in the limit of large $\gamma$, in complete agreement with the detailed 
calculation of the present paper. 

\item
The validity of the kinetic transport coefficients $\eta_{\rm K}$ and $\zeta_{\rm 
K}$ and the convergence of the Chapman-Enskog expansion require that spatial 
variations ($\mu \sim l_0 \mbox{\boldmath $\nabla$}$) are small over a 
characteristic kinetic lengthscale $l_0 \sim \bar{v} t_0 \sim \bar{v}/ n \gamma 
R^d_0$. The convergence of the gradient expansion in (\ref{dispart}) and the 
validity of the dissipative viscosities $\eta_{\rm D}$ and $\zeta_{\rm D}$ require 
in addition that spatial variations are 
small over the diameter of an action sphere, $R_0\sim \bar{v} t_w$.
Both criteria pose bounds on the shear rates, imposed in the simulations, as well as on the validity of the Chapman-Enskog expansion.

\item
An interesting extension of the present theory would be towards generalised 
hydrodynamics. Such a region exists if $t_w \gg t_0$ or $R_0\gg l_0$. Then, the 
hydrodynamic modes with wave numbers $k$ in the range $(2 \pi/R_0, 2 \pi/l_0)$ 
have $k$-dependent dissipative viscosities $\eta_{\rm D}$ and $\zeta_{\rm D}$. 
They may be calculated by studying the eigenmodes of the linearised FPB equation 
(\ref{eqn:LINEARIZED}).
A similar wavevector-range to generalised hydrodynamics occurs in
dense hard sphere fluids, where $l_0$ is small compared to the hard
sphere diameter $R_0$. Such theories have been used successfully to
describe neutron scattering experiments on liquid argon and liquid
sodium \onlinecite{Schepper}. Generalised hydrodynamics in DPD might
therefore be of interest in explaining light and neutron experiments on 
concentrated colloidal suspensions.

\item
The equilibrium properties (see Figure 2 and \onlinecite{MY}) and transport 
coefficients of DPD (see \onlinecite{GG:NEQ,BE:CORREL}) depend sensitively on the 
step size $\delta t$. The Fokker-Planck equation (\ref{eqn:FP}), the detailed 
balance condition (\ref{eqn:DB_CONSTRAINT}), the FPB equation (\ref{eqn:BBGKY}), 
the hydrodynamic equations (\ref{ns}) and corresponding transport coeffcients in 
subsections 1,2,and 3 of Section \ref{section7} only hold for the continuous time 
model $(\delta t \rightarrow 0)$. The only analytic study, available on DPD at 
finite $\delta t$ \onlinecite{MY}, calculates the equilibrium temperature 
$\theta(\delta t)$, and derives criteria, imposed on $\delta t$, for the stability 
of the equilibrium distribution $f_0({\bf x})$.

\end{enumerate}

The most important open problem on DPD is a systematic analysis of all
$\delta t$-corrections to equilibrium and transport properties, such
as an explanation of Figure 2 and 3, suggesting that the current form
of the modified DPD algorithm for {\em finite} step size $\delta t$
does not obey the detailed balance conditions, which implies that its stationary 
state is not the thermal equilibrium state described by the Gibbs' distribution.

\section*{Acknowledgements}

C.M. acknowledges financial support from EPSRC(UK), ERASMUS(EU) and Unilever
Research(UK) and the hospitality of Universiteit Utrecht where most of this work 
was carried out.
G.B. acknowledges the financial support of the foundation ``Fundamental
Onderzoek der Materie'' (FOM), which is financially supported by the
Dutch National Science Foundation (NWO).
We would also like to thank P. Espa\~{n}ol, D. Ten Bosch, J. Van den Berg, N. Van 
Kampen, T. van Noije, P. Warren and J. Yeomans for helpful discussions and 
correspondence.

\appendix
\section{Details of the Macroscopic flow equations}
\label{appendix1}

\subsection{Momentum Conservation equation}

Inserting $A=\sum_i m {\bf v}_i \delta ({\bf r}-{\bf r}_i)$ in (\ref{eqn:GRCE})
yields directly:

\begin{equation} \label{eqn:MOM1}
\partial_t (\rho {\bf u}) = -\mbox{\boldmath $\nabla$} \cdot \left< \sum_i m {\bf 
v}_i {\bf v}_i
\delta({\bf r}-{\bf r}_i) \right> + \left< \sum_i {\bf F}_i \delta
({\bf r}-{\bf r}_i) \right> - m\gamma \left< \sum_{i,j \neq i} w(R_{ij})
\left( {\bf \hat{R}}_{ij}.{\bf v}_{ij} \right) {\bf \hat{R}}_{ij} \delta
({\bf r}-{\bf r}_i) \right>.
\end{equation}
The first term on the rhs, which will be called rhs1, is transformed to the local 
rest frame of the fluid
by introducing peculiar velocities,${\bf V}_i = {\bf v}_i - {\bf u}({\bf r}_i
,t)$ and yields:

\begin{equation} \label{rhs1.1}
\mbox{rhs1} = -\mbox{\boldmath $\nabla$} \cdot (\rho {\bf uu} + {\bf \Pi}_{\rm 
K}),
\end{equation}
where

\begin{equation} \label{rhs1.2}
{\bf \Pi}_{\rm K} = \left< \sum_i m{\bf V}_i {\bf V}_i \delta({\bf r}-{\bf r}_i)
\right>
\end{equation}
is the kinetic part of the pressure tensor, as listed in (\ref{eqn:PT_CONTRIBS}).
The second term on the rhs of (\ref{eqn:MOM1}), which will be called rhs2,
involves the conservative interparticle forces ${\bf F}_i = \sum_{j \neq i}
{\bf F}({\bf R}_{ij})$.  Symmetrising over $i$ and $j$ yields then:

\begin{eqnarray} \label{rhs2.1}
\mbox{rhs2} &=& \left< \frac{1}{2} \sum_{i, j \neq i} {\bf F}({\bf R}_{ij})
\left[ \delta({\bf r}-{\bf r}_i) - \delta({\bf r}-{\bf r}_j) \right]
\right> \nonumber \\
&=& -\mbox{\boldmath $\nabla$} \cdot \left< \frac{1}{2} \sum_{i, j \neq i} {\bf 
F}({\bf R}_{ij})
{\bf R}_{ij} \int^{1}_{0} \! d\lambda \: \delta({\bf r}-{\bf r}_i
+\lambda {\bf R}_{ij} ) \right> \nonumber \\
&\equiv& -\mbox{\boldmath $\nabla$} \cdot {\bf \Pi}_{\rm C}.
\end{eqnarray}
Here we have used the identity

\begin{eqnarray} \label{eqn:IK_RESULT}
\delta({\bf r}-{\bf r}_i) - \delta({\bf r}-{\bf r}_j) &=& -\int^1_0 \!
d\lambda \: \frac{d}{d\lambda} \delta({\bf r}-{\bf r}_i+\lambda {\bf R}_{ij})
\nonumber \\
&=&-\mbox{\boldmath $\nabla$} \cdot {\bf R}_{ij} \int^1_0 \! d\lambda \: \delta 
({\bf r}-{\bf r}_i
+\lambda {\bf R}_{ij}).
\end{eqnarray}
The third term on the rhs of (\ref{eqn:MOM1}), referred to as rhs3, is due
to dissipative particle interactions and can be treated in a similar fashion.
Symmetrising over $i$ and $j$, and replacing $\delta({\bf r}-{\bf r}_i)$
by $(1/2) [\delta({\bf r}-{\bf r}_i)-\delta({\bf r}-{\bf r}_j)]$ we obtain:

\begin{eqnarray} \label{rhs3.1}
\mbox{rhs3} &=& \mbox{\boldmath $\nabla$} \cdot \left< \frac{m}{2} \sum_{i, j \neq 
i} \gamma w(R_{ij})
{\bf R}_{ij} {\bf \hat{R}}_{ij} \left( {\bf \hat{R}}_{ij}.{\bf v}_{ij}\right)
\int^1_0 \! d\lambda \: \delta( {\bf r}-{\bf r}_i + \lambda {\bf R}_{ij})
\right> \nonumber \\
&\equiv& -\mbox{\boldmath $\nabla$} \cdot {\bf \Pi}_{\rm D}.
\end{eqnarray}
The results (\ref{rhs2.1}) and (\ref{rhs3.1}) are of the same general form,
and can be expressed using the pair distribution function (\ref{eqn:F_DEFN}) as:

\begin{equation} \label{usef2}
\left< \sum_{i,j \neq i} A({\bf R}_{ij},{\bf v}_i, {\bf v}_j) \delta
({\bf r}-{\bf r}_i+\lambda {\bf R}_{ij}) \right> = \int d{\bf v} \int d{\bf v}'
\int d{\bf R} A({\bf R},{\bf v},{\bf v}') f^{(2)}({\bf v},{\bf r}+\lambda
{\bf R},{\bf v}',{\bf r}+(\lambda-1){\bf R},t).
\end{equation}
where (\ref{rhs2.1}), (\ref{rhs3.1}) and (\ref{usef2}) yield respectively
${\bf \Pi}_{\rm C}$ and ${\bf \Pi}_{\rm D}$ as listed in (\ref{eqn:PT_CONTRIBS}) 
with $\overline{f}^{(2)}$ defined in (\ref{eqn:F2BAR}).
Combination of (\ref{eqn:MOM1}), (\ref{rhs1.1}), (\ref{rhs2.1}) and (\ref{rhs3.1})
gives the macroscopic equation for the momentum density,

\begin{equation} \label{eqn:MOM_RESULT}
\partial_t (\rho {\bf u}) + \mbox{\boldmath $\nabla$} \cdot \left\{ \rho {\bf uu} 
+
{\bf \Pi}_{\rm K} + {\bf \Pi}_{\rm C} + {\bf \Pi}_{\rm D} \right\} = 0,
\end{equation}
as listed in equations (\ref{eqn:MACRO_CONS}) and (\ref{eqn:PT_DEFN}) in the
body of the paper.

\subsection{Energy balance equation}

We start with the kinetic energy density $e_{\rm K}$ by setting
$A=\sum_i (1/2) m {\bf v}_i^2 \delta ({\bf r}-{\bf r}_i)$ in (\ref{eqn:GRCE}).
This yields, after some algebra:

\begin{eqnarray} \label{kenergy}
\partial_t e_{\rm K} &=&  -\mbox{\boldmath $\nabla$} \cdot \left< \sum_i 
\frac{1}{2}
m {\bf v}_i^2 \delta({\bf r}-{\bf r}_i) \right> + \left<
\sum_{i, j \neq i} {\bf v}_i \cdot {\bf F}({\bf R}_{ij}) \delta({\bf r}
-{\bf r}_i) \right> \nonumber \\
&-& m\gamma \left< \sum_{i, j \neq i} w(R_{ij}) \left( {\bf \hat{R}}_{ij}
\cdot {\bf v}_{ij} \right) \left( {\bf \hat{R}}_{ij} \cdot {\bf v}_i \right)
\delta)({\bf r}-{\bf r}_i) \right> + \gamma \theta_0 \left<
\sum_{i,j \neq i} w(R_{ij}) \delta ({\bf r}-{\bf r}_i) \right>.
\end{eqnarray}
By setting 
$A=\frac{1}{2} \sum_{i, j \neq i} \phi(R_{ij}) \delta({\bf r}-{\bf r}_i)$
we find similarly for the potential energy density,

\begin{equation} \label{penergy}
\partial_t e_{\rm \phi} = -\mbox{\boldmath $\nabla$} \cdot \frac{1}{2} \left< 
\sum_{i, j \neq i}
{\bf v}_i \phi(R_{ij}) \delta ({\bf r}-{\bf r}_i) \right> -\frac{1}{2}
\left< \sum_{i, j \neq i} {\bf v}_{ij}.{\bf F}({\bf R}_{ij}) \delta
({\bf r}-{\bf r}_i) \right>.
\end{equation}
We sum (\ref{kenergy}) and (\ref{penergy})
to obtain the rate of change of the {\em total} energy density:

\begin{equation} \label{tenergy}
e = e_{\rm K} + e_{\rm \phi} = \left< \sum_i
\epsilon_i ({\bf v}) \delta({\bf r}-{\bf r}_i) \right>, 
\end{equation}
where $\epsilon_i({\bf v})$ is the microscopic energy per particle:

\begin{equation} \label{microenergy}
\epsilon_i({\bf v}) = \frac{1}{2} m {\bf v}_i^2 + \frac{1}{2}
\sum_{j \neq i} \phi (R_{ij}).
\end{equation}

We denote the $n$-th term on the rhs of (\ref{kenergy}) and (\ref{penergy})
by (an) and (bn) respectively and get the following results:

\begin{eqnarray} \label{energyresults}
\mbox{(a1)}+\mbox{(b1)} &=& -\mbox{\boldmath $\nabla$} \cdot \left< \sum_i {\bf 
v}_i \epsilon_i
({\bf v}) \delta({\bf r}-{\bf r}_i) \right> \equiv -\mbox{\boldmath $\nabla$} 
\cdot {\bf q}_{\rm K}
\nonumber \\
\mbox{(a2)}+\mbox{(b2)} &=& \frac{1}{2} \left< \sum_i \left( {\bf v}_i
+{\bf v}_j \right) \cdot {\bf F}({\bf R}_{ij}) \delta ({\bf r}-{\bf r}_i)
\right> \nonumber \\
&=& -\mbox{\boldmath $\nabla$} \cdot \left< \frac{1}{4} \sum_{i, j \neq i} {\bf 
R}_{ij} {\bf F}({\bf R}_{ij})
\cdot \left( {\bf v}_i+{\bf v}_j \right) \int^1_0 \! d\lambda \: \delta
({\bf r}-{\bf r}_i+\lambda {\bf R}_{ij}) \right> \nonumber \\
&=&-\mbox{\boldmath $\nabla$} \cdot {\bf q}_{\rm C}.
\end{eqnarray}
The expression for [(a2)+(b2)] has been symmetrized over $i$ and $j$ and
(\ref{eqn:IK_RESULT}) has been used.  The term (a4) represents the energy
source $\Gamma_{\rm R}$ caused by the random forces.  In (a3) we split
${\bf v}_i$ into $(1/2) {\bf v}_{ij} + (1/2) ({\bf v}_i+{\bf v}_j)$.  The first
term containing ${\bf v}_{ij}$ gives the energy sink $\Gamma_{\rm D}$ resulting
from the damping forces.  The second term containing ${\bf v}_{ij}$ is again 
symmetrized and combined with (\ref{eqn:IK_RESULT}) to give the dissipative part 
of the energy current $-\mbox{\boldmath $\nabla$} \cdot {\bf q}_{\rm D}$. 
Combination of these terms then gives:

\begin{eqnarray} \label{otherenergy}
&&\mbox{(a4)}+\mbox{(a3)} = \gamma \left< \sum_{i,j \neq i} w(R_{ij})
\delta ({\bf r}-{\bf r}_i) \right> -\frac{1}{2} m\gamma \left<
\sum_{i,j \neq i} w(R_{ij}) \left( {\bf \hat{R}}_{ij} \cdot {\bf v}_{ij}
\right)^2 \delta ({\bf r}-{\bf r}_i) \right> \nonumber \\
&&+\mbox{\boldmath $\nabla$} \cdot \left< \frac{1}{4} \sum_{i,j \neq i} {\bf 
R}_{ij} w(R_{ij})
\left( {\bf \hat{R}}_{ij} \cdot {\bf v}_{ij} \right) {\bf \hat{R}}_{ij}
\cdot \left( {\bf v}_i+{\bf v}_j \right) \int^1_0 \! d\lambda \: \delta
({\bf r}-{\bf r}_i+\lambda {\bf R}_{ij}) \right>
+\gamma \left< \sum_{i,j \neq i} w(R_{ij}) \delta ({\bf r}-{\bf r}_i) \right>
\nonumber \\
&&\equiv \Gamma_{\rm R} - \Gamma_{\rm D} - \mbox{\boldmath $\nabla$} \cdot {\bf 
q}_{\rm D}.
\end{eqnarray}
where $\Gamma_{\rm R}$, $\Gamma_{\rm D}$ and ${\bf q}_{\rm D}$ are defined by
the three preceding terms respectively.

To obtain the full energy balance equation we sum (\ref{tenergy}) to 
(\ref{otherenergy}) to obtain

\begin{eqnarray} \label{energyflow}
\partial_t e  &=& -\mbox{\boldmath $\nabla$} \cdot \left[ {\bf q}_{\rm K}
+{\bf q}_{\rm C} + {\bf q}_{\rm D} \right] + \Gamma_{\rm R} - \Gamma_{\rm D}
\nonumber \\
&\equiv& - \mbox{\boldmath $\nabla$} \cdot {\bf q} + \Gamma \mbox{.}
\end{eqnarray}


\begin{references}

\bibitem{Doolen}
G.D. Doolen, ``Lattice Gas Methods for Partial Differential Equations'' 
(Addison-Wesley Publishing Company, 1990).

\bibitem{RZ}
D.H. Rothman and S. Zaleski, Rev. Mod. Phys. {\bf 66}, 1417 (1994).


\bibitem{Succi}
R. Benzi, S. Succi and M. Vergassola, Phys. Rep. {\bf 222}, 145 (1992).

\bibitem{HK:DPD}
P.J. Hoogerbrugge and J.M.V.A. Koelman, Europhys.Lett. {\bf 19}, 155 (1992).

\bibitem{PEPW:DPD}
P. Espa\~nol and P. Warren, Europhys.Lett. {\bf 30}, 191 (1995); P. Espa\~nol, Phys.Rev.E {\bf 53}, 1572 (1996).

\bibitem{PE:HD}
P. Espa\~nol, Phys.Rev.E {\bf 52}, 1734 (1995); 

\bibitem{KH:POLY}
J.M.V.A. Koelman and P.J. Hoogerbrugge, Europhys. Lett. {\bf 21}, 363 (1993).

\bibitem{SHM:POLY}
A.G. Schlijper, P.J. Hoogerbrugge and C.W. Manke, J. Rheol. {\bf39}, 567 (1995).

\bibitem{Madden}
W.G. Madden, Y. Kong, C.M. Manke and A.G. Schlijper, Internat. Journ. 
Thermophysics {\bf 15}, 1093 (1994). 


\bibitem{CN:DPD}
P.V. Coveney and K.E. Novik, Phys.Rev.E {\bf 54}, 5134 (1996).

\bibitem{BCL}
E.S. Boek, P.V. Coveney and H.N.W. Lekkerkerker,
J.Phys.Cond.Matt. {\bf 8}, 9509 (1996); E.S. Boek, P.V. Coveney and
  H.N.W. Lekkerkerker and P. van der Schoot (preprint).

\bibitem{ME1}
C. Marsh, Internal Report Oxford University, June 1996 (unpublished).

\bibitem{VDB}
J. van den Berg, D. ten Bosch and M.H. Ernst, Master's thesis
university of Utrecht, June 1996, and Internal Shell Report
SIEP-96-5071.

\bibitem{MY}
C. Marsh and J. Yeomans, ``Dissipative particle dynamics: the equilibrium for 
finite time steps'', to be published in Europhys.Lett. 1997.

\bibitem{Groot}
S.R. de Groot and P. Mazur, ``Non-equilibrium Thermodynamics''
(North Holland Publishing Company, Amsterdam 1969).

\bibitem{GG:NEQ}
G. Grinstein, J.Appl.Phys. {\bf 69}, 5441 (1991);
D.H. Lee, G. Grinstein and S. Sachdev, Phys.Rev.Lett. {\bf
64}, 1927 (1990).


\bibitem{DKS:CORREL}
T.R. Kirkpatrick, J.R. Dorfman and J.V. Sengers,  Annu.Rev.Phys.Chem.
{\bf 45}, 213 (1994).

\bibitem{BE:CORREL}
H.J. Bussemaker and M.H. Ernst, Phys.Rev.E {\bf 53}, 5837 (1996).

\bibitem{RESIBOIS}
P. Resibois and M. de Leener, ``Classical Kinetic Theory of Liquids'' 
(Wiley, New York 1977).

\bibitem{CC}
S. Chapman and T.G. Cowling, ``The Mathematical Theory of Non-Uniform
Gases'' (Cambridge University Press, 3rd edition 1970).

\bibitem{JR}
J.T. Jenkins and M.W. Richman, Phys.Fluids {\bf 28}, 3485 (1985).

\bibitem{source}
Please distinguish the bold face ${\bf \Gamma}$, denoting a point in phase space, 
from the light type $\Gamma$, denoting sources and sinks.



\bibitem{COHEN}
E.G.D. Cohen, 
``Fundamental Problems in Statistical Mechanics'', Vol.I 
(North Holland Publishing Company, Amsterdam 1962), p. 110.

\bibitem{VK}
N.G. Van Kampen, ``Stochastic Processes in Physics and Chemistry'', Chap. 8
(North Holland Publishing Company, Amsterdam 1992).

\bibitem{Schepper}
E.G.D. Cohen, I.M. de Schepper and M.J. Zuilhof, Physica B {\bf 127}, 282 (1984); 
P. Verkerk, J. Westerweel, U. Bafile, L.A. de Graaf, W. Monfrooij and I.M. de 
Schepper, Phys. Rev. A {\bf 40}, 2860 (1989).

\bibitem{REICHL}
L.E. Reichl, ``A Modern Course in Statistical Physics'', Chap. 13
(University of Texas Press, Austin 1980).

\end{references}
\end{document}